\documentclass[entropy,article,accept,pdftex,moreauthors]{Definitions/mdpi} 
\firstpage{1} 
\pubvolume{1}
\issuenum{1}
\articlenumber{0}
\pubyear{2025}
\copyrightyear{2025}
\externaleditor{Gregg Jaeger}
\datereceived{8 March 2025} 
\daterevised{25 April 2025} 
\dateaccepted{28 April 2025} 
\datepublished{ } 
\hreflink{https://doi.org/} 

\usepackage{graphicx}           
\usepackage{bm}                 
\usepackage{amsmath}            
\usepackage{amssymb}
\usepackage{verbatim}           
\usepackage{amsthm}             

\def\bra#1{{\langle#1|}}
\def\ket#1{{|#1\rangle}}
\def\bracket#1#2{{\langle#1|#2\rangle}}

\def\tr{{\rm Tr}}

\def\x{{\hat x}}

\Title{Quantum Electrodynamics from Quantum Cellular Automata, and the Tension Between Symmetry, Locality, and Positive Energy} 

\TitleCitation{Quantum Electrodynamics from Quantum Cellular Automata, and the Tension Between Symmetry, Locality, and Positive Energy}

\Author{Todd A. Brun* 
 and Leonard Mlodinow\footnote{lmlodinow@gmail.com}}
\AuthorNames{Todd A. Brun and Leonard Mlodinow}

\AuthorCitation{Brun, T.A.; Mlodinow, L.}

\address[1]{Center for Quantum Information Science and Technology, University of Southern California, \mbox{Los Angeles, CA 90089, USA} 
}

\corres{Correspondence: tbrun@usc.edu,  Tel.: +1-213-740-3503 
}

\date{\today}

\abstract{Recent work has demonstrated a correspondence that bridges quantum information processing and high-energy physics: discrete quantum cellular automata (QCA) can, in the continuum limit, reproduce quantum field theories (QFTs). This QCA/QFT correspondence raises fundamental questions about how matter/energy, information, and the nature of spacetime are related. Here, we show that free QED is equivalent to the continuous-space-and-time limit of Fermi and Bose QCA theories on the cubic lattice derived from quantum random walks satisfying simple symmetry and unitarity conditions. In doing so, we define the Fermi and Bose theories in a unified manner using the usual fermion internal space and a boson internal space that is six-dimensional. We show that the reduction to a two-dimensional boson internal space (two helicity states arising from spin-1 plus the photon transversality condition) comes from restricting the QCA theory to positive energies. We briefly examine common symmetries of QCAs and how time-reversal symmetry demands the existence of negative-energy solutions. These solutions produce a tension in coupling the Fermi and Bose theories, in which the strong locality of QCAs seems to require a non-zero amplitude to produce negative-energy states, leading to an unphysical cascade of negative-energy particles. However, we show in a 1D model that, by extending interactions over a larger (but finite) range, it is possible to exponentially suppress the production of negative-energy particles to the point where they can be neglected.}


\begin{document}

\section{Introduction}

Quantum walks (QWs) \cite{Aharonov93,Ambainis01,Aharonov01,Kempe03} connect information theory with physical law by providing a discrete model of particle dynamics. In such systems, a quantum walk is based on a simple set of principles and symmetries, with a particle located at any of the vertices of a graph or lattice. In each time step, the particle moves to one of the neighboring vertices (connected by an edge). The particle’s time evolution is given by a unitary transformation that, in the continuum limit, can lead to a relativistic wave equation, such as the Weyl or Dirac equation, or to Maxwell’s equations \cite{BialynickiBirula94,Watrous95,Meyer96a,Meyer96,Bracken07,Chandrashekar10,DAriano14,Arrighi14,Farrelly14,MlodinowBrun18}. In recent years, this correspondence has been extended to quantum field theories (QFTs) \cite{Bisio15a,Bisio15b,Mallick16,MlodinowBrun20,BrunMlodinow20,Arrighi20,MlodinowBrun21,Eon23,Centofanti24,Bisio25}. The idea is that one can construct a quantum cellular automaton (QCA) from the quantum walk and then recover the desired Lorentz-invariant QFT in the limit of continuous time and energies. Several pieces of recent work have grappled with the issues that arise in adding interactions to QCAs \cite{Arrighi20,MlodinowBrun21,Eon23,Centofanti24}, which would be necessary to recover interacting QFTs. That is one of the major topics of this paper.

Quantum information processing comprises the action of a sequence of unitary operations (quantum gates) on an initial state of a set of quantum systems (e.g., qubits, which are the fundamental units of quantum information). In QFT, the time development of a quantum field is given by the action of a unitary operator on a state describing quantum particles or the creation and annihilation operators corresponding to them. The QCA/QFT correspondence described here suggests that systems of the latter type can arise from systems of the former type. This reveals a deep and fundamental connection between those two seemingly unrelated systems.

For example, quantum field theories are traditionally constructed by starting with a classical field and then quantizing it, while the QCA/QFT correspondence suggests that what is fundamental is not a classical field but an underlying quantum walk. This is reminiscent of Wheeler’s suggestion ``It from Bit,'' in which he proposed that information plays a significant role in the foundation of physics \cite{Wheeler89}. The universe, he suggested, may be fundamentally an information processing system from which the apparent reality of matter somehow emerges. The correspondence may also have more practical \mbox{implications---providing} a new calculational approach to simulation on a quantum computer \cite{Eon23,Farrelly20,Lloyd96}.

But there is another reason to consider the relation of discrete models to quantum field theory: the possibility that space may indeed be discrete, or at least that there is different physics at the Planck scale. Though continuous spacetime is an almost universal assumption in physics, even in classical physics, the idea of a point particle such as an electron inevitably leads to divergences. Gravity, too, does not allow an arbitrarily high concentration of gravitational charge (i.e., energy) in a small region of spacetime, because if it is too concentrated, an object will collapse to a black hole. Yet, to make observations at shorter and shorter distances, one must focus ever larger energies into a small volume. In both QFT and general relativity (GR), these short-range problems become serious---both theories are plagued by singularities and infinities at short distances. In QFT, the usual remedy is renormalization, but the methods are mathematically dubious. In GR, it is broadly accepted that, at small distance scales, the theory breaks down and is no longer applicable. As a result, some have speculated that a quantum theory of gravity will naturally lead to the idea that spacetime is not continuous, with the likely values of a fundamental unit of space and time being (at least roughly) the Planck length $\ell \approx 10^{-33}$ cm and the Planck time $t \approx 10^{-43}$ s. In this view, spacetime is composed of discrete, tiny cells, as is the case in the QCA models of field theory.

An obvious objection to discrete spacetime is that we would lose Lorentz invariance. But though the lattice QCA models do not admit Lorentz invariance, it is recovered in the limiting case of small lattice spacing (or long wavelength). Lorentz symmetry, then, could be an approximation, an effective symmetry valid at length scales larger than the fundamental length. 
Probing shorter lengths would uncover small deviations from the Dirac evolution. In particular, a lattice structure would be reflected in a modification of the free-particle dispersion relation. In a discrete universe with a fundamental length and time on the Planck scale, current technology could not distinguish between full Lorentz invariance and the approximate invariance of the discrete system. This situation may soon change, and experiments involving neutron diffraction have been proposed to detect the Lorentz non-invariance that would result if a lattice-based theory represented the true laws of nature, with the Dirac theory being only a continuous approximation \cite{BrunMlodinow19}.

In this paper, we will show how free Quantum Electrodynamics (QED) arises from quantum walks on a lattice and their corresponding QCAs. We will detail how imposing a handful of basic principles leads to the Fermi (Dirac) and Bose (Maxwell) theories and discuss how they can be coupled. We show that a difficulty in going to an interacting theory is that the symmetries of these QCAs---particularly translation and time-reversal symmetry---lead to theories with both positive- and negative-energy solutions. These solutions are highly delocalized, which means that localized couplings will in general lead to the production of negative-energy particles. We discuss this problem, analyze a one-dimensional example, and look at ways to overcome it.

In Section~\ref{sec2}, we begin by defining appropriate discrete-time quantum walks for fermions and bosons on a lattice \cite{MlodinowBrun18,MlodinowBrun21}. The walk represents the one-particle sector of the theory. Once this is defined, it can be promoted to the multi-particle case in much the same way that one defines a Fock space from a single-particle theory. In Section~\ref{sec3}, we impose the condition that the walk not have a preferred lattice axis. We will see this implies that although the dynamics will not be Lorentz-invariant for any finite lattice spacing, Lorentz invariance will be recovered in the long-wavelength limit. In Section~\ref{sec4}, we diagonalize the physical space (but not the internal space) factor in the evolution operator by changing to the momentum representation of the wave function. For low-momentum fermions with mass, and bosons that are massless, we take the interval $\Delta t$ to be small and define an approximate time derivative operator, allowing us to obtain the momentum-space representation of the Dirac and Maxwell equations. In Section~\ref{sec5}, we show that it is possible to promote the single-particle fermionic and bosonic theories to multi-particle QCAs that reduce to the usual Dirac and Maxwell field theories in the long-wavelength limit. In \mbox{Section~\ref{sec6}, we} discuss symmetries of the lattice theories, along with their roles, especially the relation of time-reversal symmetry to negative-energy states. In Section~\ref{sec7}, we discuss issues in the inclusion of interactions between the Fermi and Bose theories. Section~\ref{sec8} presents a short summary of our results.

\section{Structure of the Walks}\label{sec2}

In deriving quantum field theories of fermions and bosons from quantum cellular automata, the first step is to define an appropriate discrete-time quantum walk on a \mbox{lattice \cite{BialynickiBirula94,Bracken07,Chandrashekar10,DAriano14,Arrighi14,Farrelly14,MlodinowBrun18,MlodinowBrun21}.} The walk represents the one-particle sector of the theory. Once this is defined, it can be promoted to the multi-particle case in much the same way that one defines a Fock space from a single-particle theory.

The quantum walk is a unitary analog of a classical random walk. In a classical random walk, in each time step, the particle has some probability of moving to the neighboring vertices. In a quantum walk, the evolution is unitary, and the particle can move to a superposition of positions.

We can think of the possible particle positions as being the vertices of a graph, with edges between neighboring vertices. For the quantum walks in this paper, the graph is always regular, 
with every vertex having the same number of neighbors. The number of neighbors is the degree of the graph, labeled by $d$. 
We can label each of the $d$ outgoing edges from a vertex $1,\ldots,d$, with that label describing a direction the particle can move in. To start, we will assume that this graph is finite, but we will later consider the infinite limit. 

To maintain both unitarity and nontrivial dynamics, the particle generally must have an internal degree of freedom, a ``coin'' space. The Hilbert space has the form \mbox{$\mathcal{H}_{\rm QW} = \mathcal{H}_X \otimes \mathcal{H}_C$,} where $\mathcal{H}_X$ is the Hilbert space of the particle position, and $\mathcal{H}_C$ is the coin space. As we will see, for fermions, the coin space has even dimension, and the coin space operators that appear in the time evolution are operators of half-integer spin. For bosons, the corresponding operators are rotation group generators with integer spin.

The graph we consider is the body-centered cubic lattice, consisting of a set of vertices $\mathbf{x} \equiv (x,y,z) = (i\Delta x, j\Delta x, k\Delta x)$, where $i, j, k$ are integers and $\Delta x$ is a fixed lattice spacing. Thus, two vertices are neighbors if one is (x,y,z) and the other is $(x\pm \Delta x, y\pm \Delta x, z\pm \Delta x)$. A discrete-time walk on this lattice involves moving one unit $\Delta x$ in each of the three cardinal directions at each time step. In the standard formulation of a quantum walk in three spatial dimensions, the evolution from one time to the next is given by a unitary evolution operator $U_{\rm QW}$ \cite{BialynickiBirula94} of \mbox{the form} 
\[
\ket{\psi_{t+\Delta t}} = U_{\rm QW} \ket{\psi_t},
\]
where
\begin{equation}
U_{\rm QW} \ket{\psi_t} = \left( \sum_j S_j \otimes A_j \right) \left(I\otimes C\right) \ket{\psi_t}.
\end{equation}
The $\{S_j\}$ are shift operators that move the particle from its current position to its neighbor in the direction $j$, and the operators $A_j$ and the unitary $C$ act on an internal space. $C$ scrambles the faces, so that one does not constantly move in the same direction. The overall unitarity of the walk is not guaranteed: unitarity requires that the operators $\{A_j\}$, which act on an internal space, satisfy the condition \cite{BialynickiBirula94,DAriano14}
\begin{equation}
U_{\rm QW}U_{\rm QW}^\dagger = \sum_{j,j'} S_j S^\dagger_{j'} \otimes A_j A^\dagger_{j'} = I.
\end{equation} 

In this paper, we use a different formulation \cite{MlodinowBrun18,Chandrashekar11,Chandrashekar13}. We decompose each time step into three successive moves, one in each cardinal direction:
\begin{equation}
U_{\rm QW} = e^{i\theta Q} \left(S_X P^+_X + S^\dagger_X P^-_X \right) \left(S_Y P^+_Y + S^\dagger_Y P^-_Y \right) \left(S_Z P^+_Z + S^\dagger_Z P^-_Z \right)
\label{eq:fermionQW}
\end{equation}
for fermions, and
\begin{equation}
U_{\rm QW} = e^{i\theta Q} \left(S_X P^+_X + P^0_X + S^\dagger_X P^-_X \right) \left(S_Y P^+_Y + P^0_Y + S^\dagger_Y P^-_Y \right) \left(S_Z P^+_Z + P^0_Z + S^\dagger_Z P^-_Z \right)
\label{eq:bosonQW}
\end{equation}
for bosons.

In the above, if the particle is a fermion, at each time step, it will move to a neighboring position in each direction. If it is a boson, there is the additional possibility that it remains in place when one or two of the direction factors are applied, but not all three (which would correspond to nothing happening and thus can be excised from our description). Thus, for fermions, the $P_j^\pm$ operators are three pairs of orthogonal projectors acting on the coin space, each satisfying $P_j^+ + P_j^-  = I$. For bosons, the $P_j^\pm$ and $P_j^0$ operators are three triplets of orthogonal projectors acting on the coin space, each satisfying $P_j^+ + P_j^- + P_j^0  = I$. 

The unitary coin flip operator is $C(\theta) = e^{-i\theta Q}$, where $Q$ is a Hermitian operator that acts on the internal space. $Q$ flips the forward and backward directions, i.e., $Q P_j^+ Q = P_j^-$. For bosons, we also have $Q P_j^0 Q = P_j^0$. We can, without loss of generality, assume that $Q^2 = I$ and $\tr\{Q\} = 0$. Note that if $U_{\rm QW}$ is parity-invariant, the equations we require for $Q$ are obeyed by the parity transformation, $\mathcal{P}$: $\mathcal{P}\ket{\mathbf{x}} = \ket{-\mathbf{x}}$ implies that $\mathcal{P} P_j^+ \mathcal{P} = P_j^-$, $\mathcal{P} P_j^- \mathcal{P} = P_j^+$. 

The idea is that the quantum walk proceeds in a manner analogous to the classical walk, through a series of coin flips. The projectors $P_j^\pm$ (and $P_j^0$ in the case of bosons) correspond to different faces of the coin, which indicate in which direction to move; the unitary $C$ scrambles the faces, so that one does not constantly move in the same direction. Unlike classical random walks, this evolution is always invertible: interference effects play a profound role in the time development of the system.

This formulation is inherently unitary. For fermions, the $\gamma$ matrices (and the Dirac equation as the long-wavelength limit) arise from the condition that the different walk axes and directions be uncorrelated (as we explain below). We obtain an analogous result for bosons, leading to the Maxwell equations.

\section{Internal Space Operators from Symmetry Requirements}\label{sec3}

The Dirac and Maxwell equations are Lorentz-invariant, but one cannot demand rotational invariance of a walk on a discrete lattice. Here, we impose the condition that the walk not have a preferred lattice axis. We will see that although the dynamics will not be Lorentz-invariant for any finite lattice spacing, Lorentz invariance will be recovered in the long-wavelength limit.

By ``not having a preferred axis'', we mean that it should be equally likely for the walker to move in any of the eight allowed directions. This is equivalent to demanding that movement along the different axes is, generically, not correlated. A move in the positive $X$ direction, for example, should not bias the move in the positive or negative $Y$ or $Z$ directions. To achieve this, we impose a noncorrelation condition on the 3D walk, which we call the {\it equal norm condition}. 

To formulate the equal norm condition for fermions, we begin by noting that shifts in the positive or negative direction along an axis, $j$, are determined by the projectors $P_j^+$ or $P_j^-$, respectively \cite{Aharonov93,Ambainis01,Aharonov01,Kempe03,MlodinowBrun18}. If we require $\tr\{P_j^+\} = \tr\{P_j^-\}$, the walk will not be biased in the positive or negative direction. We also do not want the shifts along different axes to be correlated. We achieve that by requiring that any eigenstate of $P_i^\pm$ have equal amplitude in the $P_j^+$ and $P_j^-$ subspaces, for all $i \ne j = X, Y, Z$. This requirement is equivalent to the following condition on the projectors:
\begin{equation}
P_i^k P_j^+ P_i^k = P_i^k P_j^- P_i^k = \frac{1}{2} P_i^k ,
\end{equation}
where $k=\pm$, $i,j=X,Y,Z$, and $i\ne j$.

If we introduce operators the $\Delta P_j = P_j^+ - P_j^-$, we can see that this equal norm condition is equivalent to requiring that the operators $\Delta P_j$ all anti-commute with each other. Given that, it is easy to see that $\{Q,\Delta P_j \} = 0$ as well. Four mutually anti-commuting operators require a coin space of dimension $D \ge 4$, leading us to identify
\begin{equation}
Q = \gamma_0, \ \ \Delta P_X = \gamma_0\gamma_1, \ \ \Delta P_Y = \gamma_0\gamma_2, \ \ \Delta P_Z = \gamma_0\gamma_3 ,
\end{equation}
where the $\{\gamma_i\}$ are the gamma matrices from the Dirac equation. In other words, we have
\begin{equation}
P_j^+ = \frac{1}{2}\left( I + \gamma_0 \gamma_j \right), \ \ P_j^- = \frac{1}{2}\left( I - \gamma_0 \gamma_j \right) .
\end{equation}

The equal norm condition has a simple geometric interpretation. There are two projection operators associated with each dimension, corresponding to the forward and backward directions, and they project onto orthogonal spaces. If the internal space is four-dimensional, for example, each of those spaces is two-dimensional, so they project onto orthogonal planes within the four-dimensional internal space. Similarly, the pairs associated with the other two space dimensions project onto orthogonal planes. However, the planes associated with $X$-dimension projection operators are not orthogonal to the planes associated with $Y$- and $Z$-dimension projectors. The equal norm condition requires that all those non-orthogonal planes be oriented at equal angles to each other. It implies that the form of the quantum walk is preserved by rotations that map the lattice axes into each other. Also, if we change the order in which we apply the steps in the $X$, $Y$, and $Z$ directions, we will obtain another walk of the same form, with the same properties, and this walk will have the same continuum limit.

Because bosons have a different internal space, allowing for the possibility that the walker remain in place, the equal norm condition for bosons is different \cite{MlodinowBrun21}. As before, 
\begin{equation}
P_i^k P_j^+ P_i^k = P_i^k P_j^- P_i^k = c P_i^k ,
\end{equation}
where $k=\pm$, $i,j=X,Y,Z$, $i\ne j$, and $c$ is some positive real constant.

With regard to the $P_j^0$ operators, we also want the amplitude for the particle to remain in its initial location without moving to be zero. This condition requires that $P_X^0 P_Y^0 P_Z^0 = 0$, which implies that either $P_X^0 P_Y^0 = 0$ or $P_Y^0 P_Z^0 = 0$ or both. Because we want to treat motion along all three axes similarly and not have the form of the evolution operator be highly dependent on the order of the shifts, we choose the symmetric condition
\begin{equation}
P_X^0 P_Y^0 = P_Y^0 P_Z^0 = P_Z^0 P_X^0 = 0 .
\end{equation}

Finally, we require an equal norm condition for the $P_j^0$ that is analogous to that above, for the $P_j^\pm$: 
\begin{equation}
P_i^k P_j^0 P_i^k = c' P_i^k ,\ \ P_i^0 P_j^k P_i^0 = c' P_i^0 ,
\end{equation}
where, again, $k=\pm$, $i,j=X,Y,Z$, $i\ne j$, and $c'$ is some positive real constant.

Again, there is a simple geometric interpretation, analogous to that in the fermion case. Here, the $P_j^0$, $P_j^+$ and $P_j^-$ are one-dimensional projectors; each projects onto some vector $\ket{k}_j$, where $k=0,+,-$ and $j=X,Y, Z$. The above conditions tell us the following:
\begin{enumerate}[labelsep=14pt]
\item For fixed $j = X, Y, Z$, the three vectors $\ket{k}_j$ are orthogonal. 
\item The vectors $\ket{0}_j$ and $\ket{0}_{j'}$ with $j \ne j'$ are orthogonal.
\item The inner products $_j\bracket{k}{k'}_{j'}$ for $k,k'=+,-$ and $j' \ne j$ all have equal magnitude.
\item The inner products$_j\bracket{0}{k}_{j'}$ for $k=+,-$ and $j' \ne j$ all have equal magnitude.
\end{enumerate}
These conditions are met if we choose  
\begin{equation}
P^\pm_{X,Y,Z} = \frac{1}{2}\left( J^2_{X,Y,Z} \pm J_{X,Y,Z} \right) ,\ \ {\rm and}\ \ 
P^0_{X,Y,Z} = I - J^2_{X,Y,Z} ,
\end{equation}
where the $J$ matrices form a spin-1 representation of $O(3)$:
\begin{equation}
J_X = \left(\begin{array}{ccc} 0 & 0 & 0 \\ 0 & 0 & -i \\ 0 & i & 0 \end{array}\right) ,\ \ 
J_Y = \left(\begin{array}{ccc} 0 & 0 & i \\ 0 & 0 & 0 \\ -i & 0 & 0 \end{array}\right) ,\ \ 
J_Z = \left(\begin{array}{ccc} 0 & -i & 0 \\ i & 0 & 0 \\ 0 & 0 & 0 \end{array}\right) .
\label{eq:jMatrices}
\end{equation}
This gives us $\Delta P_j = J_j$. 

Note that $J_i$ plays the role of $\gamma_0 \gamma_i$ in the above fermion expression for $\Delta P_{X,Y,Z}$. For fermions, in the 1D and 2D cases as well as the massless 3D case, only a two-dimensional internal space is necessary, with $\Delta P_i = \sigma_i$. For the 3D massive fermion case, the dimension of the internal space must be doubled, with $\Delta P_i = \sigma_i \rightarrow \Delta P_i = \sigma\otimes\sigma_i = \gamma_0 \gamma_i$, where the choice of the first $\sigma$ depends on the representation (it is $\sigma_X$ for the Dirac representation, with $\gamma_0 = \sigma_Z \otimes 1$). The doubling is due to the number of anti-commuting operators required and leads to a theory accommodating two spin states as well as particles of both matter and antimatter.

In our theory, the internal space of the boson theory must also be doubled. As we will see, when we calculate the dynamics, the model just described already includes anti-particle/negative-energy states (as well as longitudinal zero-energy states). Since the photon is its own anti-particle, the negative-energy sector does not carry additional information, and we will restrict ourselves to the positive-energy sector of the free theory. We will see, however, that the dynamics of the above theory ties the polarization to the sign of the energy: right-handed photons come with a positive energy and left-handed photons with a negative energy, and there is no parity operator. To see the absence of a parity operator, 
note that the parity operator $\mathcal{P}$ would satisfy $\mathcal{P} \Delta P_i \mathcal{P} = - \Delta P_i$, which is not possible for all three operators, \mbox{i = 1, 2, 3 (see Equation~(\ref{eq:jMatrices})).} A doubling of the internal space is thus necessary to allow the inclusion of both transverse polarizations, as well as a parity operator. We will therefore work with a six-dimensional internal space. Versions of Maxwell’s equations with six-dimensional field vectors have been studied in other contexts (see, for example, Ref.~\cite{Mohr}).

To double the internal space, we proceed along the lines of the fermion case:
\[
J_i \rightarrow \sigma \otimes J_i  ,
\]
where we will choose $\sigma= \sigma_Z$ to give a Dirac-like representation. With this substitution, the projection operators become
\begin{eqnarray}
P_j^0 &=& 1 \otimes (1 - J_j^2) , \nonumber\\
P_j^+ &=& \frac{1}{2} (1 \otimes J_j^2 + \sigma_Z \otimes J_j) , \nonumber\\
P_j^- &=& \frac{1}{2} (1 \otimes J_j^2 - \sigma_Z \otimes J_j) ,
\end{eqnarray}
and $\Delta P_j = \sigma_Z \otimes J_j$. 

We can now define a parity operator $\mathcal{P} = \sigma_X \otimes 1$, which takes $P_j^+$ into $P_j^-$ and vice versa. This could be taken as the coin operator $Q = \mathcal{P}$, as we did in the fermion case. However, as we will see, the coin generator endows the particle with mass; so, since we are interested in (massless) photons here, we will set $Q = 0$. 

To preserve a common structure for both fermions and bosons, it is useful to define boson gamma matrices in analogy with the usual fermion ones:
\begin{eqnarray}
\gamma_0^{\rm bose} &\equiv&  \sigma_X\otimes 1 = \mathcal{P} , \nonumber\\
\gamma_i^{\rm bose} &\equiv& -i \sigma_Y\otimes J_i .
\end{eqnarray}
With these choices, $\Delta P_i = \sigma_Z \otimes J_i = \gamma_0^{\rm bose} \gamma_i^{\rm bose}$. 

We note from the discussion of time-reversal symmetry in Section~\ref{sec6} that the minimum dimension of the internal space that would give two positive-energy solutions is four. However, we know of no construction of a QW with a four-dimensional internal space that gives rise to the Maxwell theory in the long-wavelength limit. So, we proceed with our six-dimensional construction, at the cost of including zero-energy solutions that decouple from the dynamics.

\section{Momentum-Space Picture and the Dirac and Maxwell Equations}\label{sec4}

We can diagonalize the physical space factor in the evolution operator (though not the factor that operates on the internal space) by changing to the momentum representation of the wave function. If the position states are $\ket{x,y,z}$, where $x = j\Delta x$, $y = k\Delta x$, and $z = l\Delta x$ for some integers $j, k, l$, then the momentum states are
\begin{equation}
\ket{k_X, k_Y k_Z} = \frac{1}{\mathcal{N}} \sum_{j,k,l} e^{-i \left(k_X j\Delta x + k_Y k\Delta x + k_Z l \Delta x \right)} \ket{j\Delta x,k\Delta x,l\Delta x} ,
\end{equation}
where $-\pi/\Delta x < k_{X,Y,Z} \le \pi/\Delta x$. For mathematical simplicity, we assume that the overall size of the lattice is finite, albeit large, with periodic boundary conditions; this choice makes the momentum discrete and the momentum states renormalizable. If the cubic lattice is $N\times N\times N$, then the normalization factor is $\mathcal{N} = N^{3/2}$. If necessary, we can then go to a limit where $N\rightarrow\infty$. In the finite case, the momentum values $k_{X,Y,Z}$ are integer multiples of $2\pi/N\Delta x$.

The states $\ket{k_X, k_Y k_Z}$ are eigenstates of the shift operators $S_{X,Y,Z}$ with eigenvalues $e^{i k_{X,Y,Z}\Delta x}$. In this representation, the evolution operator (with $Q = \gamma_0$ for fermions and $Q=0$ for photons) is
\begin{eqnarray}
U &=& e^{-i\theta Q} e^{i K_X\Delta x\Delta P_X} e^{i K_Y\Delta x\Delta P_Y} 
e^{i K_Z\Delta x\Delta P_Z} \nonumber\\
&=& \sum_{\mathbf{k}} \ket{\mathbf{k}}\bra{\mathbf{k}} \otimes U_{\mathbf{k}}
\equiv \sum_{\mathbf{k}} \ket{\mathbf{k}}\bra{\mathbf{k}} \otimes V_{\mathbf{k}} \Lambda_{\mathbf{k}} V^\dagger_{\mathbf{k}} ,
\label{eq:evolutionMomentum}
\end{eqnarray}
where $K_{X,Y,Z}$ are the operators corresponding to the components of the momentum, and $\{\Lambda_{\mathbf{k}}\}$ are the diagonalized forms of $\{U_{\mathbf{k}}\}$. Using the choices above for $\Delta P_{X,Y,Z}$, we have
\begin{equation}
U_{\mathbf{k}} = e^{-i\theta \gamma_0} e^{i k_X\Delta x\gamma_0\gamma_1}
e^{i k_Y\Delta x\gamma_0\gamma_2}  e^{i k_Z\Delta x \gamma_0\gamma_3},
\end{equation}
where, for fermions, these are the usual gamma matrices, and for bosons, the gamma matrices are the boson versions defined, and the first term is absent, i.e., $\theta = 0$. (As we will see, this corresponds to the massless case.) In this representation, we see that the Hilbert space decomposes into $D$-dimensional blocks ($D = 4$ for fermions, $D=6$ for bosons) labeled by the momentum $\mathbf{k}$. The energy eigenstates $\ket{\mathbf{k},\lambda_{\mathbf{k},j}}$ of the quantum walk can be found by diagonalizing the $D\times D$ matrices $U_{\mathbf{k}} = V_{\mathbf{k}} \Lambda_{\mathbf{k}} V^\dagger_{\mathbf{k}}$. Since $U_{\mathbf{k}}$ is unitary, the matrices $\Lambda_{\mathbf{k}}$ have eigenvalues of the form $\lambda_{\mathbf{k},j} = e^{i\phi_{\mathbf{k},j}}$.

For fermions, if we go to the long-wavelength/low-mass limit $|\mathbf{k}\Delta x| \ll 1$ and $|\theta| \ll 1$, we obtain
\begin{equation}
U_{\mathbf{k}} \approx 1 + i\Delta x \left(k_X \Delta P_X + k_Y \Delta P_Y + k_Z \Delta P_Z \right) - i \theta Q .
\end{equation}

For low-momentum particles with mass, we can take the interval $\Delta t$ to be small and define an approximate time derivative operator:
\begin{equation}
\partial_t \equiv \frac{U-1}{\Delta t} .
\end{equation}

In terms of this, we can obtain a Dirac equation, which holds in the long-wavelength/low-mass limit:
\begin{equation}
\partial_t \ket\psi = i \frac{\Delta x}{\Delta t} \left(k_X \Delta P_X + k_Y \Delta P_Y + k_Z \Delta P_Z \right)\ket\psi - i \frac{\theta}{\Delta t} Q\ket\psi .
\end{equation}

We can relate the physical space interval $\Delta x$ and the physical time interval $\Delta t$ in terms of the speed of light $c$: $c \equiv \Delta x/\Delta t$. Then, defining the particle mass $\theta/\Delta t \equiv m_e c^2$ and using $\Delta P_i = \gamma_0 \gamma_i$, we obtain the momentum-space representation of the Dirac equation:
\begin{equation}
\gamma_0 \partial_t \ket\psi = i c \left(k_X \gamma_1 + k_Y \gamma_2 + k_Z \gamma_3 - m_e c \gamma_0 \right)\ket\psi .
\end{equation}

For bosons, we obtain the analogous equation, except with $\Delta P_i = \gamma_0^{\rm bose} \gamma_i^{\rm bose}$. This gives, in the same small $\mathbf{k}$ limit,
\begin{equation}
U_{\mathbf{k}} \approx 1 + i\Delta x \left(k_X \gamma^{\rm bose}_0 \gamma^{\rm bose}_1 + k_Y \gamma^{\rm bose}_0 \gamma^{\rm bose}_2 + k_Z \gamma^{\rm bose}_0 \gamma^{\rm bose}_3 \right) .
\end{equation} 
We thus obtain 
\begin{eqnarray}
\partial_t &\equiv& \frac{U-1}{\Delta t} \nonumber\\
&=& i c \left(k_X \gamma^{\rm bose}_0 \gamma^{\rm bose}_1 + k_Y \gamma^{\rm bose}_0 \gamma^{\rm bose}_2 + k_Z \gamma^{\rm bose}_0 \gamma^{\rm bose}_3 \right) \nonumber\\
&=& i c \gamma^{\rm bose}_0 \left( \mathbf{k} \cdot \pmb{\gamma}^{\rm bose} \right)
= i c \sigma_Z \otimes \left( \mathbf{k} \cdot \mathbf{J} \right) .
\end{eqnarray}

We can define an effective Hamiltonian
\begin{equation}
H \equiv c \sigma_Z \otimes \left( \mathbf{k} \cdot \mathbf{J} \right)
= i c \sigma_Z \otimes \left( \begin{array}{ccc} 0 & -k_Z & k_Y \\ k_Z & 0 & - k_X \\
- k_Y & k_X & 0 \end{array} \right) .
\end{equation} 
Using the identity
\begin{equation}
\left(\mathbf{k} \cdot \mathbf{J}\right) \mathbf{v} = i \mathbf{k}\times\mathbf{v} ,
\label{eq:vecidentity}
\end{equation}
we see that the eigenvectors of $\left( \mathbf{k} \cdot \mathbf{J} \right)$ are
\begin{equation}
\mathbf{v}_0 = \frac{1}{k} \left( \begin{array}{c} k_X \\ k_Y \\ k_Z \end{array}\right) ,\ \ 
\mathbf{v}_\pm = \frac{1}{\sqrt{2k^2\left( k_X^2+k_Z^2\right)}}
\left( \begin{array}{c} - i k k_Z \pm k_X k_Y \\ \mp \left( k_X^2+k_Z^2\right) \\ i k k_X \pm k_Y k_Z \end{array}\right) ,
\end{equation}
with eigenvalues $0$ and $\pm ik$, respectively, where $k=|\mathbf{k}|=\sqrt{k_X^2 + k_Y^2 + k_Z^2}$. These correspond to a longitudinal mode, and right- and left-polarized modes, respectively. The Hamiltonian thus has two zero-energy longitudinal eigenstates, as well as a positive- and negative-energy eigenstate for each circular polarization. In the free theory, these do not mix, so if we consider positive-energy initial states, the dynamics do not take us out of the positive-energy subspace. That is not generally true in the presence of interactions, which we will discuss later.

When we promote the single-particle theory to allow for the creation and annihilation of photons, we will find that the field operators, expressed in the usual manner in terms of the creation and annihilation operators and the above eigenvectors, will satisfy the Maxwell equations in the long-wavelength limit. To see that here, for the one-particle sector, we have to consider the photon wave function  (see Refs.~\cite{BialynickiBirula,Darwin32}):
\begin{equation}
\pmb{\Psi} = \left( \begin{array}{c} \pmb{\Psi_+} \\ \pmb{\Psi_-} \end{array}\right) ,
\end{equation}
where $\pmb{\Psi_+} = (\pmb{E} + i\pmb{B})$,  and $\pmb{\Psi_-} = \pmb{\Psi_+}^*$. Using the Schr\"odinger equation
\begin{equation}
i\partial_t \pmb{\Psi} = H \pmb{\Psi} ,
\end{equation}
and the vector identity in Equation~(\ref{eq:vecidentity}), we obtain
\begin{equation}
i\partial_t \left(\begin{array}{c} \pmb{\Psi_+} \\ \pmb{\Psi_-} \end{array}\right) =
\left(\begin{array}{c} c\mathbf{k}\times \pmb{\Psi_+} \\ -c\mathbf{k}\times \pmb{\Psi_-} \end{array}\right) .
\end{equation} 
Substituting $-i\pmb{\nabla}$ for $c\mathbf{k}$ and considering the real and imaginary parts separately, we obtain the two time-dependent Maxwell equations:
\begin{eqnarray}
\partial_t \mathbf{E} &=& \pmb{\nabla}\times\mathbf{B} ,\nonumber\\
\partial_t \mathbf{B} &=& -\pmb{\nabla}\times\mathbf{E} .
\end{eqnarray} 
The other two Maxwell equations (in the absence of charge and current) arise from the transversality of the positive-energy eigenvectors.

\section{Promotion to Quantum Cellular Automata}\label{sec5}

\subsection{Fermions}

As was shown in \cite{MlodinowBrun21}, it is possible to promote the single-particle fermionic theory to a multi-particle QCA that reduces to the usual Dirac field theory in the long-wavelength limit. To accomplish this, one begins with a maximum number of particles, $N_{\rm max}$, and then lets $N_{\rm max}$ become arbitrarily large. In other words, one begins with the space 
\begin{equation}
\mathcal{H}_{\rm total} = \mathcal{H}^{(1)} \otimes \cdots \otimes \mathcal{H}^{(N_{\rm max})} ,
\end{equation}
where $\mathcal{H}^{(j)} = \mathcal{H}_{\rm QW} \oplus {\rm span}(\ket{\omega})$ is the Hilbert space for particle type $j$, which contains either one quantum walk particle or no particle (the vacuum state $\ket{\omega}$). The system then evolves by the unitary 
\begin{equation}
U_{\rm total} = U^{(1)} \otimes \cdots \otimes U^{(N_{\rm max})} ,
\end{equation}
where
\begin{equation}
U^{(j)} = U_{\rm QW} \oplus \ket\omega \bra\omega .
\end{equation} 

We then proceed with essentially the usual Fock space construction, restricting ourselves to the totally antisymmetric subspace of $\mathcal{H}_{\rm total}$. We call this the physical subspace, 
which we define as
\begin{equation}
\mathcal{H}_{\rm phys} = \mathcal{H}_0 \oplus \mathcal{A}_1 \oplus \cdots \oplus \mathcal{A}_{N_{\rm max}} .
\end{equation} 
The subspace $\mathcal{A}_n$ is the totally antisymmetric subspace with $n$ particles, and $\mathcal{H}_0$ is the one-dimensional vacuum space ${\rm span}(\ket\Omega)$, where $\ket\Omega = {\ket\omega}^{\otimes N_{\rm max}}$. We can define basis vectors for the subspaces $\mathcal{A}_n$ in terms of the ``energy eigenstates'' of the quantum walk:
\begin{eqnarray}
\ket\Omega &=& {\ket\omega}^{\otimes N_{\rm max}} , \nonumber\\
\ket{\mathbf{k},j} &=& \ket{\mathbf{k},\lambda_{\mathbf{k},j}}
\otimes {\ket\omega}^{\otimes N_{\rm max}-1} , \\
\ket{\mathbf{k}_1,j_1; \mathbf{k}_2,j_2} &=& \frac{1}{\sqrt2}
\bigl( \ket{\mathbf{k}_1,\lambda_{\mathbf{k}_1,j_1}}
\otimes \ket{\mathbf{k}_2,\lambda_{\mathbf{k}_2,j_2}} \nonumber\\
&& - \ket{\mathbf{k}_2,\lambda_{\mathbf{k}_2,j_2}}
\otimes \ket{\mathbf{k}_1,\lambda_{\mathbf{k}_1,j_1}} \bigr)
\otimes {\ket\omega}^{\otimes N_{\rm max}-1} , \nonumber
\end{eqnarray}
and so forth, defining basis states $\ket{\mathbf{k}_1,j_1; \ldots; \mathbf{k}_n,j_n}$ for particle numbers $n$ up to $N_{\rm max}$. The state $\ket{\mathbf{k}_1,j_1; \ldots; \mathbf{k}_n,j_n}$ is an eigenstate of the evolution operator $U_{\rm total}$ with eigenvalue
\[
\lambda = \exp\left( i \sum_{\ell=1}^n \phi_{\mathbf{k}_\ell, j_\ell} \right) .
\]

The $n$-particle antisymmetrized subspace is thus preserved by the time evolution, and, hence, so is the full physical subspace $\mathcal{H}_{\rm phys}$: an antisymmetric initial state will evolve to be antisymmetric at all later times. Given this basis, we can define a set of creation and annihilation operators that transform the energy basis vectors into each other:
\begin{equation}
\ket{\mathbf{k}_1,j_1; \ldots; \mathbf{k}_n,j_n} \equiv
a^\dagger_{\mathbf{k}_1,j_1} \cdots a^\dagger_{\mathbf{k}_n,j_n} \ket\Omega ,
\end{equation}
and that obey the usual anti-commutation relations. To avoid redundant counting of the basis vectors, we should establish a standard ordering for the states $\mathbf{k}_i,j_i$. The choice of this ordering is unimportant, so long as it is unambiguous.

We can write the action of the local evolution U in terms of these operators for any $\ket\Psi \in \mathcal{H}_{\rm phys}$:
\begin{equation}
U_{\rm total} \ket\Psi = \exp\left( i \sum_{\mathbf{k},j} \phi_{\mathbf{k},j} a^\dagger_{\mathbf{k},j} a_{\mathbf{k},j} \right) \ket\Psi .
\label{eq:antisymmetricU}
\end{equation}

A very important observation should be made here. These creation and annihilation operators $a^\dagger_{\mathbf{k},j}$ and $a_{\mathbf{k},j}$ are clearly {\it nonlocal} operators; so the unitary operator on the right-hand side of Equation~(\ref{eq:antisymmetricU}) is nonlocal. But the unitary operator $U_{\rm total}$ is itself purely local. It is a collection of local quantum walk unitaries and can easily be embedded in a quantum cellular automaton. These two sides are equal only on states in the physical subspace. Indeed, the right-hand side has not been defined for states outside that subspace. 

Having defined creation and annihilation operators for the energy eigenstates, we can then define creation and annihilation operators $\tilde{a}_{\mathbf{k},j}^\dagger$ and $\tilde{a}_{\mathbf{k},j}$ in the momentum representation. We can write these as operator-valued column vectors $\mathbf{\tilde{a}}_{\mathbf{k},j}^\dagger$ and $\mathbf{\tilde{a}}_{\mathbf{k},j}$, where
\begin{equation}
\mathbf{\tilde{a}}_{\mathbf{k},j}^\dagger = \left(\begin{array}{c} \tilde{a}_{\mathbf{k},1}^\dagger \\
\vdots \\ \tilde{a}_{\mathbf{k},D}^\dagger \end{array}\right)
= V^\dagger_{\mathbf{k}} \left(\begin{array}{c} {a}_{\mathbf{k},1}^\dagger \\
\vdots \\ {a}_{\mathbf{k},D}^\dagger \end{array}\right) ,
\end{equation}
using the matrices $V^\dagger_{\mathbf{k}}$ from Equation~(\ref{eq:evolutionMomentum}). These vectors have an effective time evolution
\begin{equation}
U_{\rm total} \mathbf{\tilde{a}}_{\mathbf{k},j}^\dagger U_{\rm total}^\dagger
= U_{\mathbf{k}} \mathbf{\tilde{a}}_{\mathbf{k},j}^\dagger ,\ \ 
U_{\rm total} \mathbf{\tilde{a}}_{\mathbf{k},j} U_{\rm total}^\dagger
= U^*_{\mathbf{k}} \mathbf{\tilde{a}}_{\mathbf{k},j} ,
\end{equation}
where $U_{\mathbf{k}}$ is given by Equation~(\ref{eq:evolutionMomentum}).

What about the maximum number of particles $N_{\rm max}$? This may seem like an unphysical feature of the theory. In the free theory, particle number is conserved, so a solution will generally be confined to a single antisymmetric subspace $\mathcal{A}_n$. However, in the interacting case, this conservation need not hold.

One way to handle this is to start with a vacuum corresponding to a Dirac Sea, in which all the negative-energy states are occupied, while all positive-energy states are vacant. Interactions do not create or destroy particles but may move a particle from a negative- to a positive-energy state, leaving a hole (corresponding to particle--anti-particle pair creation), or move a positive-energy state to a negative-energy hole (corresponding to particle--anti-particle annihilation). In such a construction, we would have $N_{\rm max} = 2N^3$, where $N$ is the lattice size. This means that the Hilbert spaces of the local subsystems at each lattice site are enormously high-dimensional (albeit finite), but the states of the theory remain restricted to the much lower-dimensional physical subspace. In the full space, the dynamics of the free theory are fully local; the effective dynamics in the physical subspace obey Fermi statistics.

\subsection{Bosons}

We can make a similar construction for free bosons by taking $N_{\rm max}$ copies of a bosonic QW, representing distinguishable particles, and then confining ourselves to the completely symmetric subspace. So long as the total number of bosonic particles $N$ is less than $N_{\rm max}$, this system will behave like a bosonic field theory with $N$ excitations, and we can introduce creation and annihilation operators and rewrite the evolution unitary in terms of them, just as we did in the fermionic case.

This construction is sufficient for the free theory, where particle number is conserved. But it is unsatisfactory for an interacting theory, in which bosons may be created or destroyed. This issue is less critical for fermions, where no more than one particle may occupy any given state and we can introduce a Dirac Sea as the vacuum state. There is no analogous construction for bosons, since any number of bosons may occupy any state.

However, the restrictions on bosonic QCAs are also much weaker than for fermions: since bosonic creation and annihilation operators for different modes commute, these operators can be made local, and there is not a ``no-go'' theorem as there is for fermionic QCAs \cite{MlodinowBrun20}. It is therefore possible to construct bosonic QCAs in a different way, which avoids these locality issues, by using excitations of local modes as our particles. We will lay out this construction of a free bosonic QCA theory in this subsection. The key idea is quite simple:  each possible basis state (comprising a position and an internal state $\ket{\mathbf{x},\varepsilon}$) of the single particle in the quantum walk becomes a harmonic oscillator mode in the QCA. The unitary evolution operators will then trade excitations between neighboring modes.

Recall that the evolution unitary for the massless bosonic quantum walk was a product of terms of the form
\begin{equation}
 \left(S_j P^+_j + P^0_j + S^\dagger_j P^-_j \right) ,
\end{equation}
where $j=X,Y,Z$. As discussed in Section~\ref{sec4}, we can recover a Maxwell field if we use a doubled space, where each of the projectors is two-dimensional, for an internal space dimension of six. To simplify the derivation of the corresponding  QCA, we will first consider a non-doubled space, with an internal dimension three, so we can write the projectors as outer products of the eigenstates of the $J_j$ matrix:
\[
P^+_j = \ket{+}_j\bra{+} ,\ \ \ P^-_j = \ket{-}_j\bra{-} ,\ \ \ P^0_j = \ket{0}_j\bra{0} .
\]
Once we have derived the form of the QCA in this case, we can double the internal space, which is straightforward. 

Instead of using shift operators, we can rewrite this unitary factor as a product of swaps between states at neighboring sites. Let $\mathbf{\hat{j}}$ be the unit vector in the $j=X,Y,Z$ direction. The term in the evolution unitary can be written as
\begin{eqnarray}
\left(S_j P^+_j + P^0_j + S^\dagger_j P^-_j \right) &=& \sum_{\mathbf{x}}
\biggl( \ket{\mathbf{x}+\Delta x\mathbf{\hat{j}}}\bra{\mathbf{x}} \otimes  \ket{+}_j\bra{+} \nonumber\\
&& \ \ \ \ \  + \ket{\mathbf{x}-\Delta x\mathbf{\hat{j}}}\bra{\mathbf{x}} \otimes  \ket{-}_j\bra{-} 
+ \ket{\mathbf{x}}\bra{\mathbf{x}} \otimes  \ket{0}_j\bra{0} \biggr) \nonumber\\
 &=& \sum_{\mathbf{x}} \ket{\mathbf{x}}\bra{\mathbf{x}} \otimes \biggl( \ket{+}_j\bra{-}
 + \ket{-}_j\bra{+} + \ket{0}_j\bra{0} \biggr) \nonumber\\
&& \times \sum_{\mathbf{x}} \biggl( \ket{\mathbf{x}+\Delta x \mathbf{\hat{j}},-}_j \bra{\mathbf{x},+}  \nonumber\\
&& \ \ \ \ \ \ + \ket{\mathbf{x},+}_j \bra{\mathbf{x}+\Delta x \mathbf{\hat{j}},-}
+ \ket{\mathbf{x},0}_j \bra{\mathbf{x},0} \biggr) .
 \end{eqnarray} 
In other words, the shifts are performed in two steps: the $+$ and $-$ states at neighboring sites are interchanged, and then at each site, $+$ and $-$ are switched back. (The 0 states remain unchanged throughout.)

The advantage of breaking this down into two steps is that each step is a purely local operation that only affects two neighboring sites. This makes it very easy to turn into a QCA that is definitely local. In constructing the QCA, each basis state $\ket{\mathbf{x},\varepsilon}$ of the QW corresponds to a harmonic oscillator mode with creation and annihilation operators $a^\dagger_{\mathbf{x},\varepsilon}, a_{\mathbf{x},\varepsilon}$. Then, the evolution unitary becomes
\begin{eqnarray}
\hat{U}_j &=&
 \left( \bigotimes_{\mathbf{x}} \exp\left\{ -\frac{i\pi}{2}\left( a^\dagger_{\mathbf{x},+} a_{\mathbf{x},-}
 + a^\dagger_{\mathbf{x},-} a_{\mathbf{x},+} \right) \right\}  \right) \nonumber\\
&& \times \left( \bigotimes_{\mathbf{x}} \exp\left\{ -\frac{i\pi}{2}\left(
a^\dagger_{\mathbf{x}+\Delta x\mathbf{\hat{j}},-} a_{\mathbf{x},+}
 + a^\dagger_{\mathbf{x},+} a_{\mathbf{x}+\Delta x\mathbf{\hat{j}},-} \right) \right\}  \right) , \nonumber\\
 &=& \exp\left\{ -\frac{i\pi}{2}\sum_{\mathbf{x}} \left( a^\dagger_{\mathbf{x},+} a_{\mathbf{x},-}
 + a^\dagger_{\mathbf{x},-} a_{\mathbf{x},+} \right) \right\} \nonumber\\
&& \times \exp\left\{ -\frac{i\pi}{2} \sum_{\mathbf{x}} \left(
a^\dagger_{\mathbf{x}+\Delta x\mathbf{\hat{j}},-} a_{\mathbf{x},+}
 + a^\dagger_{\mathbf{x},+} a_{\mathbf{x}+\Delta x\mathbf{\hat{j}},-} \right) \right\} .
 \label{eq:bosonicQCAfactor}
\end{eqnarray}

We can see that $\hat{U}_j$ is a product of tensor products of unitaries, which act on the local subsystems $\mathbf{x}$ or on non-overlapping sets of modes at neighboring subsets $\mathbf{x}$ and $\mathbf{x}+\Delta x\mathbf{\hat{j}}$. So $\hat{U}_j$ is local and defines a QCA. Since operators on different modes commute, we can rewrite each tensor product of exponentials as the exponential of a sum of terms (where now we understand the creation and annihilation operators as acting as the identity on other modes). Further, we see that the defined QCA preserves the particle number (or the number of excitations); and on the single-particle sector, it exactly reproduces a quantum walk.

To reproduce the full QCA, of course, we need not just a single operator $\hat{U}_j$, but the product of operators producing shifts in all three directions:
\begin{equation}
\hat{U}_{\rm QCA} = \hat{U}_X \hat{U}_Y \hat{U}_Z .
\end{equation} 
Each of the operators $\hat{U}_{X,Y,Z}$ has the same form as in Equation~(\ref{eq:bosonicQCAfactor}). In the quantum walk, we obtain nontrivial evolution because the projectors $P^{\pm}_{X,Y,Z}$ for the shifts along each axis project onto subspaces that are non-orthogonal to those of the other axes. For the QCA, this means that the creation and annihilation operators in Equation~(\ref{eq:bosonicQCAfactor}) for one axis $j$ are linear combinations of those for a different axis $j'$.

Since all the expressions inside the exponentials are quadratic in creation and annihilation operators, we can rewrite Equation~(\ref{eq:bosonicQCAfactor}) in terms of a standard set of modes with operators $a^\dagger_{\mathbf{x},\ell}$ and $a_{\mathbf{x},\ell}$:
\begin{eqnarray}
\hat{U}_j &=& \exp\left\{ - i \sum_{\mathbf{x},\ell,\ell'} (m^j_{\ell \ell'} + (m^j_{\ell' \ell})^*)
\left( a^\dagger_{\mathbf{x},\ell} a_{\mathbf{x},\ell'} \right) \right\} \nonumber\\
&& \times\exp\left\{ - i \sum_{\mathbf{x},\ell,\ell'}
\left( m^j_{\ell \ell'} a^\dagger_{\mathbf{x}+\Delta x\mathbf{\hat{j}},\ell} a_{\mathbf{x},\ell'} 
+ (m^j_{\ell \ell'})^* a^\dagger_{\mathbf{x},\ell'}  a_{\mathbf{x}+\Delta x\mathbf{\hat{j}},\ell} \right) \right\} ,
\label{eq:bosonicQCAfactor2}
\end{eqnarray}
where $[m^j_{\ell \ell'}]$ are the matrix elements of the matrix $(\pi/2)\ket{-}_j\bra{+}$ in the standard basis.

One can go to the momentum picture by introducing a new set of modes corresponding to different momentum states:
\begin{equation}
a_{\mathbf{x},\ell} = \frac{1}{N^{3/2}} \sum_{\mathbf{k}} e^{-i\mathbf{x}\cdot\mathbf{k}} a_{\mathbf{k},\ell} ,\ \ \ 
a^\dagger_{\mathbf{x},\ell} = \frac{1}{N^{3/2}} \sum_{\mathbf{k}} e^{i\mathbf{x}\cdot\mathbf{k}} a^\dagger_{\mathbf{k},\ell} .
\end{equation} 
It is easy to check that creation and annihilation operators for different values of $\mathbf{k}$ commute with each other. If we substitute this definition into Equation~(\ref{eq:bosonicQCAfactor2}), the unitary decomposes into a product of factors acting on different momentum modes:
\begin{eqnarray}
\hat{U}_j &=& \prod_{\mathbf{k}} \exp\left\{ - i \sum_{\ell,\ell'} (m^j_{\ell \ell'} + (m^j_{\ell' \ell})^*)
\left( a^\dagger_{\mathbf{k},\ell} a_{\mathbf{k},\ell'} \right) \right\} \nonumber\\
&& \times\exp\left\{ - i \sum_{\ell,\ell'}
\left( e^{i k_j\Delta x} m^j_{\ell \ell'} a^\dagger_{\mathbf{k},\ell} a_{\mathbf{k},\ell'} 
+ e^{-i k_j\Delta x} (m^j_{\ell \ell'})^* a^\dagger_{\mathbf{k},\ell'}  a_{\mathbf{k},\ell} \right) \right\} ,
\label{eq:bosonicQCAfactor3}
\end{eqnarray} 
The coupling between neighboring sites in the position picture becomes a phase of $e^{i k_j\Delta x}$ in the momentum picture.

A product of unitaries produced by quadratic Hamiltonians is equivalent to a single unitary produced by a quadratic Hamiltonian. That is, given a set of field modes with annihilation operators $a_1,\ldots,a_n$ and two $n\times n$ Hermitian matrices $A$ and $B$, then there exists an $n\times n$ Hermitian matrix $C$ such that
\begin{equation}
\exp\left\{ -i \mathbf{a}^\dagger \cdot C \mathbf{a} \right\} =
  \exp\left\{ -i \mathbf{a}^\dagger \cdot B \mathbf{a} \right\} \times
  \exp\left\{ -i \mathbf{a}^\dagger \cdot A \mathbf{a} \right\} ,
\end{equation}
where
\begin{equation}
\mathbf{a} = \left( \begin{array}{c} a_1 \\ \vdots \\ a_n \end{array} \right) ,\ \ \ 
\mathbf{a}^\dagger = \left( \begin{array}{c} a_1^\dagger \\ \vdots \\ a_n^\dagger \end{array} \right) .
\end{equation} 
One can calculate the matrix $C$ from the matrices $A$ and $B$ by solving the equation
\begin{equation}
\exp\left\{ i C^T \right\} =  \exp\left\{ i B^T \right\} \exp\left\{ i A^T \right\} .
\end{equation} 
This is most readily achieved by diagonalizing $A$ and $B$, exponentiating them, and then multiplying them. One can find $C$ by diagonalizing the resulting matrix and taking its log. In the case of the unitary factor $U_j$, the matrices $A_j$ and $B_j$ have elements
\begin{eqnarray}
a_{\ell \ell'} &=& e^{i k_j\Delta x} m^j_{\ell \ell'} + e^{-i k_j\Delta x} (m^j_{\ell' \ell} )^* , \nonumber\\
b_{\ell \ell'} &=& m^j_{\ell \ell'} + (m^j_{\ell' \ell} )^* .
\end{eqnarray}

The result of this series of three unitaries is a quadratic evolution with a matrix $C$, where
\begin{equation}
\exp\left\{ i C^T \right\} =  \exp\left\{ i B_x^T \right\} \exp\left\{ i A_x^T \right\} 
\exp\left\{ i B_y^T \right\} \exp\left\{ i A_y^T \right\} 
\exp\left\{ i B_z^T \right\} \exp\left\{ i A_z^T \right\} .
\end{equation} 
These matrices are $3\times3$, so, in general, it is not difficult to diagonalize them. For a bosonic QW, we suggested in Section~\ref{sec3} that the basis vectors should be eigenvectors of the three spin-1 $J$ matrices given in Equation~(\ref{eq:jMatrices}). This choice for the bosonic QCA would lead to three pairs of $A$ and $B$ matrices:
\begin{eqnarray}
e^{iA^T_x} = \left(\begin{array}{ccc} 1 & 0 & 0 \\ 0 & - i\cos(k_X\Delta x) & i\sin(k_X\Delta x) \\
0 & i\sin(k_X\Delta x) & i\cos(k_X\Delta x) \end{array}\right) , &&
e^{iB^T_x} = \left(\begin{array}{ccc} 1 & 0 & 0 \\ 0 & -i & 0 \\
0 & 0 & i \end{array}\right) ,\nonumber\\
e^{iA^T_y} = \left(\begin{array}{ccc} i\cos(k_Y\Delta x) & 0 & - i\sin(k_Y\Delta x) \\ 0 & 1 & 0 \\
- i\sin(k_Y\Delta x) & 0 & -i\cos(k_Y\Delta x) \end{array}\right) , &&
e^{iB^T_y} = \left(\begin{array}{ccc} i & 0 & 0 \\ 0 & 1 & 0 \\
0 & 0 & -i \end{array}\right) , \\
e^{iA^T_z} = \left(\begin{array}{ccc} - i\cos(k_Z\Delta x) & i\sin(k_Z\Delta x) & 0 \\
i\sin(k_Z\Delta x) & i\cos(k_Z\Delta x) & 0 \\ 0 & 0 & 1 \end{array}\right) , &&
e^{iB^T_z} = \left(\begin{array}{ccc} -i & 0 & 0 \\
0 & i & 0 \\ 0 & 0 & 1 \end{array}\right) . \nonumber
\end{eqnarray} 
Multiplying out these matrices gives us
\begin{equation}
\exp\{iC^T\} = \left(\begin{array}{ccc} c_y c_z & -c_y s_z & s_y \\
c_x s_z + s_x s_y c_z & c_x c_z - s_x s_y s_z & -s_x c_y \\
s_x s_z - c_x s_y c_z & s_x c_z + c_x s_y s_z & c_x c_y \end{array}\right) ,
\end{equation}
where we have used the compact notation $c_{x,y,z} = \cos(k_{x,y,z}\Delta x)$ and $s_{x,y,z} = \sin(k_{x,y,z}\Delta x)$. 

This matrix can, in principle, be diagonalized in closed form, but the expression is dauntingly complicated. Its eigenvalues can also be calculated, though they are also somewhat complicated:
\begin{eqnarray}
\lambda_0 &=& 1 = e^0 , \nonumber\\
\lambda_+ &=& \frac{1}{2}\left( c_x c_y + c_y c_z + c_z c_x - s_x s_y s_z - 1 \right)  \nonumber\\
 && - i \sqrt{1 - \frac{1}{4}\left(c_x c_y + c_y c_z + c_z c_x - s_x s_y s_z - 1 \right)^2 } 
   \equiv e^{-i\phi_+} , \\
\lambda_- &=&  \frac{1}{2}\left(c_x c_y + c_y c_z + c_z c_x - s_x s_y s_z - 1 \right)  \nonumber\\
 && + i \sqrt{1 - \frac{1}{4}\left(c_x c_y + c_y c_z + c_z c_x - s_x s_y s_z - 1 \right)^2 } 
    \equiv e^{-i\phi_-} . \nonumber
\end{eqnarray}

Simple observation shows that the $+$ and $-$ eigenvalues have opposite phases: \mbox{$\phi_- = -\phi_+$.} These can be thought of as negative- and positive-energy modes. We can recover more familiar expressions by going to the long-wavelength limit, where $|\mathbf{k}\Delta x| \ll 1$. In this limit,
\begin{equation}
\exp\{iC^T\} \approx I - i( k_X J_X + k_Y J_Y + k_Z J_Z )\Delta x \approx \exp\{ - i \mathbf{k}\cdot\mathbf{J}\Delta x\} .
\end{equation} 
In this limit, the phases corresponding to the eigenvalues are approximately $\phi_0 = 0$, $\phi_+ \approx k\Delta x = \Delta x \sqrt{k_X^2 + k_Y^2 + k_Z^2}$, $\phi_- \approx - k\Delta x$. These correspond to a zero-energy, positive-energy, and negative-energy mode.

As discussed in Section~\ref{sec4}, we can recover the Maxwell field theory by doubling the internal space. In place of the spin matrices $J_j$ we use the $6\times6$ matrices $\sigma_Z \otimes J_j$. The derivation above is essentially doubled, giving two zero-energy, two negative-energy, and two positive-energy modes for each allowed momentum $\mathbf{k}$. If we confine ourselves to the positive-energy subspace, these two modes correspond to the two allowed helicity states of the electromagnetic field. The two negative-energy modes are like retarded versions of these two modes. The zero-energy modes are static and unaffected by the other modes.

We have seen that the free theories of both fermionic and bosonic QCAs include matching positive- and negative-energy modes (and zero-energy modes as well, in the bosonic case). This turns out to be a general property of QCAs based on their overall symmetry, as discussed in the next section.

\section{Symmetries of QCAs}\label{sec6}

Symmetries of QCAs---especially translation and parity symmetry---have been discussed extensively elsewhere \cite{Watrous95,Meyer96a,Meyer96}, so here we will just briefly summarize their effects on the types of solutions that exist for the systems we consider in this paper.

\subsection{Translation Symmetry}

Translation symmetry implies that the evolution operator is invariant under shifts in the spatial lattice. In a 3D cubic lattice of size $N$ with periodic boundaries, we can define shift operators $S_{X,Y,Z}$ that move the states of the local subsystems to the site $\Delta x$ away in the positive $X$, $Y$, or $Z$ direction, respectively, wrapping around at the periodic boundaries. $S^\dagger_{X,Y,Z}$ obviously represents a shift in the negative direction. Translation symmetry was extensively studied in the early QCA literature \cite{Watrous95,Meyer96a,Meyer96}. On the cubic lattice, these shifts all commute.

If $S$ is a product of any number of these unit shifts $S_{X,Y,Z}$ or $S^\dagger_{X,Y,Z}$, then the translation symmetry of the QCA evolution unitary $\hat{U}$ implies
\begin{equation}
S\hat{U}S^\dagger = \hat{U} \ \Rightarrow \ S\hat{U} = \hat{U}S \ \Rightarrow \ [\hat{U},S] = 0 .
\end{equation} 
Since $\hat{U}$ and $S$ commute, we can find a basis of eigenstates of $\hat{U}$ that are eigenstates of $S$. This is what allows the momentum picture used both for QWs and QCAs in this paper.

\subsection{Rotation}

By contrast, the discrete lattice of the QCA means that it does not have rotation symmetry: waves propagating along the axes may behave differently than waves between axes. However, QCAs derived using the equal norm condition described in \mbox{Section~\ref{sec3} recover} rotation symmetry (and indeed, Lorentz invariance) in the long-wavelength limit \cite{MlodinowBrun18}. So these continuous symmetries may arise effectively at long wavelengths.

\subsection{Parity}

Parity symmetry implies that the evolution unitary is invariant under spatial inversion, $\mathbf{x} = (x,y,z) \rightarrow -\mathbf{x} = (-x,-y,-z)$. For this to be possible, we must define the parity operator $\mathcal{P}$ to also act appropriately on the internal space. This is most straightforward to see in the QW case \cite{MlodinowBrun18}. If we have $\mathcal{P} U \mathcal{P} = U$, then
\begin{eqnarray}
U_{QW} &=& e^{i\theta Q} \prod_{j=X,Y,Z} \left( S_j P^+_j + S^\dagger_j P^-_j \right) \nonumber\\
&=& \mathcal{P} e^{i\theta Q} \prod_{j=X,Y,Z} \left( S_j P^+_j + S^\dagger_j P^-_j \right) \mathcal{P} \nonumber\\
&=&  e^{i\theta \mathcal{P}Q\mathcal{P}} \prod_{j=X,Y,Z} \left( \mathcal{P}S_j\mathcal{P}\mathcal{P} P^+_j\mathcal{P} + \mathcal{P}S^\dagger_j\mathcal{P} \mathcal{P}P^-_j\mathcal{P} \right) .
\end{eqnarray} 
Spatial inversion means that $\mathcal{P}S_j\mathcal{P} = S^\dagger_j$, which implies that
\begin{equation}
\mathcal{P}Q\mathcal{P} = Q ,\ \ \mathcal{P}P^\pm_j\mathcal{P} = P_j^{\mp} .
\end{equation}

The simplest way of satisfying these conditions is if $\mathcal{P}$ acts on the internal space like $Q$ and if $Q$ anti-commutes with the operators $\Delta P_j = P_j^+ - P_j^-$. This anti-commutation plays an important role in the emergence of the Dirac equation in the long-wavelength limit of QWs \cite{MlodinowBrun18,MlodinowBrun20,BrunMlodinow20,MlodinowBrun21}.

\subsection{Time-Reversal Symmetry and Negative Energies}

Time-reversal symmetry captures the idea that time evolution looks the same forward and backward. Classically, this would be illustrated by reversing the momenta of all particles while leaving their positions unchanged; this is like ``running the movie backwards.'' In quantum mechanics, we also need to flip the spins of particles. This symmetry is very important to the tension between locality of interactions on the one hand and positive energy on the other, as we shall see in Section~\ref{sec7}.

We can write this symmetry in terms of a time-reversal operation $\mathcal{T}$:
\begin{equation}
\hat{U} = \mathcal{T} \hat{U} \mathcal{T}^{-1} .
\label{eq:timereversal1}
\end{equation} 
Unlike the previous two examples, this operation $\mathcal{T}$ does not correspond to a unitary transformation, but to an {\it antiunitary} transformation, which includes complex conjugation of the state:
\begin{equation}
\mathcal{T} = VK ,\ \ \ \mathcal{T}^{-1} = KV^\dagger ,
\end{equation}
where $K$ is complex conjugation ($K\ket\psi = \ket\psi^*$), and $V$ is a unitary operator. Plugging this form into Equation~(\ref{eq:timereversal1}), we obtain
\begin{equation}
\hat{U} = \mathcal{T} \hat{U} \mathcal{T}^{-1} = V K\hat{U}K V^\dagger \Rightarrow
V^\dagger \hat{U} V = \hat{U}^* .
\label{eq:timereversal2}
\end{equation}

Suppose that the evolution unitary $\hat{U}$ has eigenstates $\{\ket{\phi_m}\}$ with eigenvalues $\{e^{i\phi_m}\}$. Then, $\hat{U}^*$ is also unitary, with eigenstates $\{\ket{\phi_m}^*\}$ and eigenvalues $\{e^{-i\phi_m}\}$. On the other hand, conjugating a unitary leaves its eigenvalues unchanged, so $V^\dagger\hat{U}V$ has the same eigenvalues $\{e^{i\phi_m}\}$ as $\hat{U}$. The only way a unitary $\hat{U}$ can satisfy Equation~(\ref{eq:timereversal2}), therefore, is if for every eigenvalue $e^{i\phi_m}$, it must also have a corresponding eigenvalue $e^{-i\phi_m}$. That is, every ``positive-energy'' solution must have a corresponding ``negative-energy'' solution, which leads to the existence of anti-particles. (Zero-energy solutions, if any, can be taken to themselves or each other.)

To see why $\mathcal{T}$ is interpreted as time reversal, we can put this condition in a more intuitive form. The inverse operator $\hat{U}^\dagger$ has the same eigenvectors $\{\ket{\phi_m}\}$ as $\hat{U}$ but the same eigenvalues $\{e^{-i\phi_m}\}$ as $\hat{U}^*$. We can define a new unitary operator $\tau$
\begin{equation}
\tau = \sum_m \ket{\phi_m} \left(\bra{\phi_m}^*\right)V^\dagger .
\end{equation} 
This operator satisfies the symmetry relation
\begin{equation}
\tau \hat{U} \tau^\dagger = \hat{U}^\dagger .
\label{eq:timereversal3}
\end{equation} 
The existence of such a unitary $\tau$ in Equation~(\ref{eq:timereversal3}) implies the time-reversal symmetry defined in Equation~(\ref{eq:timereversal1}). Intuitively, $\tau$ is the equivalent of the classical time-reversal operation of flipping the momenta of all particles---it transforms the state so that it evolves ``backwards'' by transforming ``positive-energy'' states into ``negative-energy'' states and vice versa.

The QCAs defined in this paper all have this time-reversal symmetry and therefore will automatically have negative-energy solutions. We note that this conclusion holds more generally even in models that are PT-symmetric---that is, invariant under applying both a parity flip and time reversal but not under each of them separately. PT symmetry would mean {that} 
\begin{equation}
\hat{U} = \mathcal{PT} \hat{U} (\mathcal{PT})^{-1} .
\end{equation} 
Since the parity-flip operator $\mathcal{P}$ is unitary, we see that $\mathcal{PT} = \mathcal{P}VK$, and we can just substitute a new unitary operator $V\rightarrow V' \equiv \mathcal{P}V$ in Equations~(\ref{eq:timereversal2}--\ref{eq:timereversal3}). The argument goes through as before, and PT-symmetric models must therefore also have negative-energy solutions. This creates a dilemma in how to go from free theories to interacting theories without allowing negative-energy solutions to be excited.

\section{Including Interaction Terms in 1D}\label{sec7}

QCAs, as described in the earlier sections, can produce free quantum field theories in their long-wavelength (or low-energy) limit. Due to their time-reversal symmetry, these QCAs have matching negative-energy and positive-energy solutions. For free theories, the existence of negative-energy solutions has no important consequences. One can restrict the initial state to include only positive-energy states. For fermionic theories, there is the additional strategy of redefining the vacuum state to be the Dirac Sea, with all negative-energy states occupied and all positive-energy states vacant. Excitation of a positive-energy state would leave a negative-energy hole: a particle--anti-particle pair. This approach has the added advantage that the dynamics can be chosen to conserve the particle number.

Having two interacting QCAs, however, will raise an immediate issue. The eigenstates of $\hat{U}$---the negative- and positive-energy states---are delocalized states of definite momentum. If the interaction is local---as would be expected in a QCA---then (as we shall see) those local operators must include a non-zero component of negative-energy operators. For fermionic QCAs with a Dirac Sea vacuum, this is not necessarily a problem; it merely means that interactions can give rise to particle--anti-particle pairs. But bosonic QCAs have no equivalent of the Dirac Sea; the negative-energy states can have any number of excitations. So, it is difficult to avoid an interaction that produces a cascade of particle--anti-particle pairs, paid for by the emission of negative-energy bosons.

To simplify the analysis, we will consider a 1D fermionic QCA coupled to a 1D bosonic QCA. One advantage of 1D is that it is possible to construct a fermionic QCA using only low-dimensional (qubit) local subsystems \cite{MlodinowBrun20}. The tension between locality and positive energy already exists in 1D, and the solutions are easier to find and numerically investigate.

\subsection{The 1D QCAs}

\begin{figure}[H]
\includegraphics[width=5in]{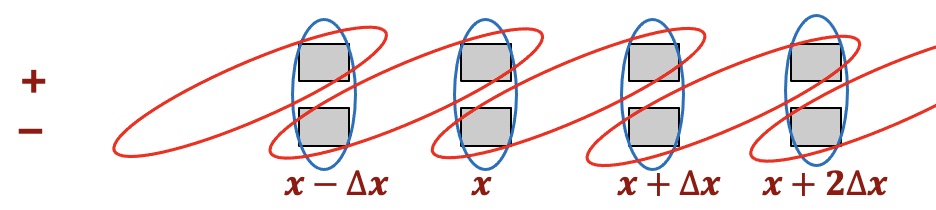}
\caption{The structure of the 1D QCAs. For the fermionic QCA, the local subsystems $x,\pm$ are qubits; for the bosonic QCA, they are harmonic oscillator modes. One step of the QCA evolution first couples the subsystems at $x,+$ and $x+\Delta x,-$ (red loops) and then the subsystems at $x,+$ and $x,-$ (blue loops).}
\label{fig:qca1D}
\end{figure}

For our 1D QCAs, each site contains two local subsystems labeled $x,+$ and $x,-$, at positions $x=j\Delta x$ for integers $0\le j<N$ with periodic boundaries. The evolution unitary alternately couples two qubits at neighboring sites and the two qubits at a single site, as shown in Figure~\ref{fig:qca1D}.
\vspace{-9pt}

In the fermionic case \cite{MlodinowBrun20}, the local subsystems are qubits, and the evolution unitary alternates two unitaries that act on pairs of local subsystems:
\begin{equation}
\hat{U}_F = \hat{C}_F \hat{\Sigma}_F .
\end{equation} 
In this equation, the coin operator is
\begin{equation}
\hat{C}_F = \cdots \otimes C_{x-\Delta x} \otimes C_x \otimes C_{x+\Delta x} \otimes \cdots ,
\end{equation}
where each of these local unitaries $C_x$ acts on the qubits at $x,+$ and $x,-$ as follows:
\begin{eqnarray}
\label{eq:coinOpDef}
C \ket{00} &=& \ket{00}, \nonumber\\
C \ket{01} &=& \cos(\theta)\ket{10} + \sin(\theta)\ket{01}, \\
C \ket{10} &=& \cos(\theta) \ket{01} - \sin(\theta)\ket{10}, \nonumber\\
C \ket{11} &=& -\ket{11}, \nonumber
\end{eqnarray}
where 1 indicates the presence (and 0 the absence) of a particle in the corresponding state, and $\theta$ is the small dimensionless parameter corresponding to the particle mass. The shift operator is
\begin{equation}
\hat{\Sigma} = \cdots \otimes S_{x-\Delta x} \otimes S_x \otimes S_{x+\Delta x} \otimes \cdots ,
\end{equation}
where each of these local unitaries $S_x$ acts on the qubits at $x,+$ and $x+\Delta x,-$ as follows:
\begin{eqnarray}
\label{eq:shiftOpDef}
S\ket{00} &=& \ket{00}, \nonumber\\
S\ket{01} &=& \ket{10}, \\
S\ket{10} &=& \ket{01}, \nonumber\\
S\ket{11} &=& -\ket{11}. \nonumber
\end{eqnarray}

In both cases, the $-$ sign on the $\ket{11}$ state produces a phase of $-1$ when two particles cross each other, producing Fermi statistics. As shown in \cite{MlodinowBrun20}, one can introduce creation and annihilation operators $b_{x,\pm}^\dagger$ and $b_{x,\pm}$ that obey the usual anti-commutation properties and rewrite the evolution unitaries in terms of them:
\begin{eqnarray}
C_x &=& \exp\left\{ - i \left(\frac{\pi}{2} + \theta\right)\left( b_{x,+}^\dagger b_{x,-} + b_{x,-}^\dagger b_{x,+} \right) \right\}, \nonumber\\
S_x &=& \exp\left\{ -\frac{i\pi}{2}\left( b_{x,+}^\dagger b_{x+\Delta x,-} + b_{x+\Delta x,-}^\dagger b_{x,+} \right) \right\} .
\label{eq:fermionQCA1D}
\end{eqnarray}

In the bosonic case, each of the local subsystems is a harmonic oscillator mode with creation and annihilation operators $a_{x,\pm}^\dagger$ and $a_{x,\pm}$. The evolution unitary $\hat{U}_B$ has the same local structure as in the fermionic QCA, but the local unitaries become
\begin{eqnarray}
C_x &=& \exp\left\{ -\frac{i\pi}{2}\left( a_{x,+}^\dagger a_{x,-} + a_{x,-}^\dagger a_{x,+} \right) \right\}, \nonumber\\
S_x &=& \exp\left\{ -\frac{i\pi}{2}\left( a_{x,+}^\dagger a_{x+\Delta x,-} + a_{x+\Delta x,-}^\dagger a_{x,+} \right) \right\} .
\label{eq:bosonQCA1D}
\end{eqnarray} 
We do not include a parameter analogous to $\theta$ in this case; these are massless bosons, which propagate in the positive or negative direction at the speed of light $c=\Delta x/\Delta t$.

We can prove that these QCAs have time-reversal symmetry by explicitly constructing the unitary operator $\tau$ in Equation~(\ref{eq:timereversal3}). Let us first make the construction in the bosonic case. Examining the unitaries defined in Equation~(\ref{eq:bosonQCA1D}), we see that every term inside the exponential includes exactly one $+$ mode and one $-$ mode. If one could, for example, flip the sign of all the operators $a_{x,+}^\dagger, a_{x,+}$, this would have the effect of flipping the sign of all the exponents, which would in turn transform $\hat{C}_B \rightarrow \hat{C}_B^\dagger$ and $\hat\Sigma_B \rightarrow \hat\Sigma_B^\dagger$. Define the \mbox{unitary operator}
\begin{equation}
V \equiv \exp\left\{ i\pi \sum_x a_{x,+}^\dagger a_{x,+} \right\} .
\end{equation} 
It is easy to check that this operator satisfies
\begin{equation}
V^\dagger a_{x,\pm} V = \mp a_{x,\pm} ,\ \ \ 
V^\dagger a_{x,\pm}^\dagger V = \mp a_{x,\pm}^\dagger .
\end{equation} 
This equation in turn implies that
\begin{equation}
V^\dagger \hat{U}_B V = V^\dagger \hat{C}_B V V^\dagger \hat\Sigma_B V = \hat{C}_B^\dagger \hat\Sigma_B^\dagger .
\end{equation} 
To complete the construction, define
\begin{equation}
\tau \equiv \hat{C}_B V^\dagger ,
\end{equation}
which satisfies
\begin{equation}
\tau \hat{U}_B \tau^\dagger = \hat{C}_B \hat{C}_B^\dagger \hat\Sigma_B^\dagger \hat{C}_B^\dagger
= \hat\Sigma_B^\dagger \hat{C}_B^\dagger = \hat{U}_B^\dagger .
\end{equation} 
An exactly analogous construction, using operators $b_{x,+}^\dagger b_{x,+}$, works for the fermionic QCA.

\subsection{Form of Interactions}

The evolution operators $\hat{U}_F$ and $\hat{U}_B$ represent free evolution of the two QCAs. For interacting QCAs, we could include an additional interaction unitary, so that the full evolution would be
\begin{equation}
\hat{U} = \hat{U}_I \hat{U}_B \hat{U}_F .
\end{equation} 
For a purely local interaction,
\begin{equation}
\hat{U}_I = \cdots \otimes U_{x-\Delta x} \otimes U_x \otimes U_{x+\Delta x} \otimes \cdots
\end{equation} 
These local unitaries $U_x$ act on the subsystems $x,\pm$ of both the bosonic and \mbox{fermionic QCAs.}

For this paper, we will assume a particular form for this interaction; but we believe that the argument regarding locality and negative energy is very general, irrespective of the particular form of the interaction. Based on the Dirac Sea construction referred to above, we choose $U_x$ to conserve the number of particles in the fermionic QCA. However, based on the examples from quantum field theory, it should not conserve the bosonic particle number. Using QFT as a model, we consider an interaction unitary of the form:
\begin{equation}
\hat{U}_I = \exp\left\{ -i \sum_x \sum_{\ell,m,n=\pm} \left( \alpha_{\ell,m,n}
a_{x,\ell} b_{x,m}^\dagger b_{x,n} + \mathrm{h.c.} \right) \right\} .
\label{eq:interaction1}
\end{equation} 
Here, the coefficients $\{ \alpha_{\ell,m,n} \}$ are parameters that represent coupling between particles in different internal states. (Generally, we will assume that these parameters are small in magnitude, $|\alpha_{\ell,m,n}| \ll 1$.) A term like the one shown in the exponential can annihilate a boson and cause a fermion to make a transition from one internal state to another; the hermitian conjugate can create a boson while causing a transition. If one thinks of states close to the Dirac Sea, where most negative-energy states are occupied and most positive-energy states are vacant, terms like this would represent a boson giving rise to a fermion--antifermion pair or a fermion--antifermion pair annihilating to produce a boson.

However, in this local picture, there is no way of knowing whether a boson produced would be positive-energy or negative-energy, and we will now see that, in general, the amplitude for both is non-zero.

\subsection{Coupling to Negative-Energy States}

For the 1D massless bosonic QCA, the ``energy'' eigenstates are just the momentum states $k,\pm$:
\begin{equation}
a_{k,\pm} = \frac{1}{\sqrt{N}} \sum_x e^{ixk} a_{x,\pm} \ \Leftrightarrow\ 
a_{x,\pm} = \frac{1}{\sqrt{N}} \sum_k e^{-ixk} a_{k,\pm} .
\label{eq:momentum1D}
\end{equation} 
The ``energies'' of $\hat{U}_B$ are $\pm \phi_k$ for $\phi_k = |k|\Delta x$. If we make the convenient choice of range for $k$ to be $-\pi/2\Delta x < k \le \pi/2\Delta x$, with $k=j\pi/N\Delta x$ for some integer $-N/2 < j \le N/2$, then the result is very intuitive: the positive-energy states are $k,+$ for $k>0$ and $k,-$ for $k\le 0$; the negative-energy states are $k,-$ for $k>0$ and $k,+$ for $k \le0$. For each $k$, there is a positive- and a negative-energy state; we can label their annihilation operators $a_{k,\mathrm{pos}}$ and $a_{k,\mathrm{neg}}$ when we want to clarify which one we mean for a general momentum $k$.

For the fermionic QCA, the ``energy'' eigenstates are also momentum eigenstates, but when $\theta\ne0$, the internal basis depends on the value of $k$. However, it will suffice just to work with momentum eigenstates for the fermions.

We can use the inverse transform from Equation~(\ref{eq:momentum1D}) to rewrite the form of the interaction unitary in Equation~(\ref{eq:interaction1}) in terms of the momentum eigenstates:
\begin{eqnarray}
\hat{U}_I &=& \exp\left\{ -\frac{i}{N^{3/2}} \sum_x \sum_{k_1,k_2,k_3} \sum_{\ell,m,n=\pm}
\left( \alpha_{\ell,m,n} e^{-i(k_1-k_2+k_3)x} a_{k_1,\ell} b_{k_2,m}^\dagger b_{k_3,n} + \mathrm{h.c.} \right) \right\} \nonumber\\
&=& \exp\left\{ -\frac{i}{N^{1/2}} \sum_{k_1,k_3} \sum_{\ell,m,n=\pm}
\left( \alpha_{\ell,m,n} a_{k_1,\ell} b_{k_1+k_3,m}^\dagger b_{k_3,n} + \mathrm{h.c.} \right) \right\} ,
\label{eq:interaction2}
\end{eqnarray}
where we derived this form by summing over $x$, which produces a Kronecker delta $N\delta_{k_2,k_1+k_3}$. From the form of this interaction, we can immediately conclude two things:
\begin{enumerate}[labelsep=14pt]
\item Momentum is conserved. The term shown in the exponent annihilates a boson with momentum $k_1$ and causes a fermion transition from a state with momentum $k_3$ to a state with momentum $k_2=k_1+k_3$. The hermitian conjugate term would cause a fermion transition from momentum $k_2$ to $k_3$ and produce a boson with momentum $k_1=k_2-k_3$.
\item The coefficients $\alpha_{\ell,m,n}$ are independent of the momentum $k$; so, if it is possible to create a positive-energy boson (say with $k>0$ and $\ell=+$), then it must also be possible to create a negative-energy boson (with $k<0$ and $\ell=+$).
\end{enumerate} 
These results are not surprising: the energy eigenstates of the boson are completely delocalized and cannot be distinguished at a single site $x$.

\subsection{Interaction Range vs. Negative-Energy Coupling}

\subsubsection{Finite-Range Interactions}

While the above argument shows that interactions localized at a single site must allow the creation of negative-energy bosons, it does not (quite) show that long-range interactions are required to prevent the production of negative-energy states. One can generalize local interactions to allow coupling between fermions and bosons over a longer, but still finite, range. A system with such finite-range interactions would still be a QCA.

To make this more concrete, consider an interaction unitary that is a product of unitaries of the following form:
\begin{equation}
\hat{U}_{I,x} = \exp\left\{ i \sum_{y=-\frac{M\Delta x}{2}}^{\frac{M\Delta x}{2}} \sum_{\ell,m,n=\pm}
\left( \alpha_{y,\ell,m,n} a_{x+y,\ell} b_{x,m}^\dagger b_{x,n} + \mathrm{h.c.} \right) \right\} ,
\label{eq:interactionrange}
\end{equation}
where $M$ is an even integer and $M/2$ is the range of the interaction. Since the coefficients $\alpha_{y,\ell,m,n}$ now depend on the relative position $y$, when we transform to the momentum picture, the transformed coefficients can depend on momentum. Is it possible to eliminate coupling to negative-energy states?

Let us pull out the bosonic piece of the interaction. A finite-range interaction as in Equation~(\ref{eq:interactionrange}) involves bosonic operators of the form
\begin{equation}
O_x = \sum_{y=-\frac{M\Delta x}{2}}^{\frac{M\Delta x}{2}} \left(
\alpha_{y,+} a_{x+y,+} + \alpha_{y,-} a_{x+y,-} \right) ,
\end{equation}
where we have fixed the internal states $m,n$ of the fermions and suppressed those indices. If we are allowed to choose the coefficients $\{\alpha_{y,\pm}\}$ arbitrarily, how small can we make the component of negative-energy operators in $O_y$ relative to the positive-energy operators? Writing the operators $a_{x+y,\pm}$ in terms of momentum operators, we obtain
\begin{equation}
O_x \equiv \sum_k \left( \tilde{\alpha}_{k,\mathrm{pos}} a_{k,\mathrm{pos}} + \tilde{\alpha}_{k,\mathrm{neg}} a_{k,\mathrm{neg}} \right) .
\end{equation}

We can calculate the coefficients $\{\tilde{\alpha}_{k,\mathrm{neg}}\}$ by rewriting the original expression for $O_x$ in terms of momentum operators:
\begin{equation}
\tilde{\alpha}_{k,\mathrm{neg}} = \frac{1}{\sqrt{N}} \sum_y \left\{ \begin{array}{cc}
e^{-i(x+y)k} \alpha_{y,-} , & k>0, \\ e^{-i(x+y)k} \alpha_{y,+} , & k\le0 .
\end{array} \right. 
\end{equation} 
We want to choose the coefficients $\{\alpha_{y,\pm}\}$ to minimize $\{\tilde{\alpha}_{k,\mathrm{neg}}\}$. We can quantify this by the sum
\begin{eqnarray}
\sum_k |\tilde{\alpha}_{k,\mathrm{neg}}|^2 &=& \frac{1}{N}\sum_{y,y'} \left[
\alpha^*_{y,-} \alpha_{y',-} \sum_{k>0} e^{-i(y-y')k} +
\alpha^*_{y,+} \alpha_{y',+} \sum_{k\le0} e^{-i(y-y')k} \right] \\
&\equiv& \sum_{j,j'=1}^{M+1} \left( \alpha^*_{(j-M/2)\Delta x,-} \alpha_{(j'-M/2)\Delta x,-} d_{jj'} +
\alpha^*_{(j-M/2)\Delta x,+} \alpha_{(j'-M/2)\Delta x,+} d^*_{jj'} \right) , \nonumber
\end{eqnarray}
where $\{d_{jj'}\}$ are elements of an $(M+1)\times(M+1)$ matrix $D$, and we have changed from summing over $y$ to summing over an integer index $j=1,\ldots,M+1$, where \mbox{$y=(j-M/2)\Delta x$.} Note that this sum is independent of $x$ (which reflects the translation symmetry of the interaction) and takes the form
\begin{equation}
\sum_k |\tilde{\alpha}_{k,\mathrm{neg}}|^2 = v_-^\dagger D v_- +  v_+^\dagger D^* v_+ ,
\end{equation}
where $D$ is the matrix with elements $d_{jj'}$ that we have just defined, and $v_\pm$ are vectors whose $M+1$ elements are the coefficients $\alpha_{y,\pm}$. This sum is minimized by choosing $v_-$ and $v_+$ to be the eigenvectors of $D$ and $D^*$, respectively, with the smallest eigenvalue. The eigenvalues of $D$ and $D^*$ are equal, and they must all be $\ge 0$, so we minimize the sum by choosing the eigenvalue closest to zero.

These matrix elements $d_{jj'}$ are
\begin{eqnarray}
d_{jj'} &=& \frac{1}{N} \sum_{\ell = 1}^{N/2} e^{-2i\pi(j-j')\ell/N} 
= \frac{e^{-2i\pi(j-j')/N}}{N} \frac{1-e^{-i\pi(j-j')}}{1-e^{-2i\pi(j-j')/N}} \nonumber\\
&=& e^{-i\pi(j-j')(N+1)/2N} \frac{\sin(\pi(j-j')/2)}{N\sin(\pi(j-j')/N)} \nonumber\\
&\approx& e^{-i\pi(j-j')/2} \frac{\sin(\pi(j-j')/2)}{\pi(j-j')} , \ \ \ N \gg M .
\end{eqnarray}  
We can make a few observations about this matrix $D$:
\begin{enumerate}[labelsep=14pt]
\item $D$ is a Toeplitz matrix: its elements $d_{jj'}$ depend only on the difference $j-j'$, so they are constant along the diagonals.
\item $D$ must be a positive matrix, and we will see that its eigenvalues lie between 0 and 1.
\item The diagonal elements $j=j'$ are $d_{jj} = 1/2$.
\item The phase factor $e^{-i\pi(j-j')/4}$ has no effect on the eigenvalues of $D$, since that phase can be absorbed into the eigenvectors $v_-$. We can instead consider the real matrix $\bar{D}$.
\end{enumerate}
\begin{equation}
\bar{d}_{jj'} = \frac{\sin(\pi(j-j')/2)}{N\sin(\pi(j-j')/N)} \approx \frac{\sin(\pi(j-j')/2)}{\pi(j-j')} .
\label{eq:realmatrix}
\end{equation} 
The numerator takes repeating values $0,1,0,-1,0,1,\ldots$, so except for $j=j'$, the even diagonals all vanish, and the odd diagonals take values of $1/\pi$, $-1/3\pi$, $1/5\pi$, $-1/7\pi$, and so forth. Interestingly, these are the coefficients of the Taylor expansion of $(1/\pi)\arctan(1)$.

For example, for $M=4$, the matrix $\bar{D}$ is
\begin{equation}
\bar{D} = \frac{1}{\pi} \left(\begin{array}{ccccc} \pi/2 & 1 & 0 & -1/3 & 0 \\ 1 & \pi/2 & 1 & 0 & -1/3 \\
0 & 1 & \pi/2 & 1 & 0 \\ -1/3 & 0 & 1 & \pi/2 & 1 \\ 0 & -1/3 & 0 & 1 & \pi/2 \end{array}\right) .
\end{equation}

\subsubsection{Toeplitz Matrices and Their Eigenvalues}

\begin{figure}[H]
\includegraphics[width=4in]{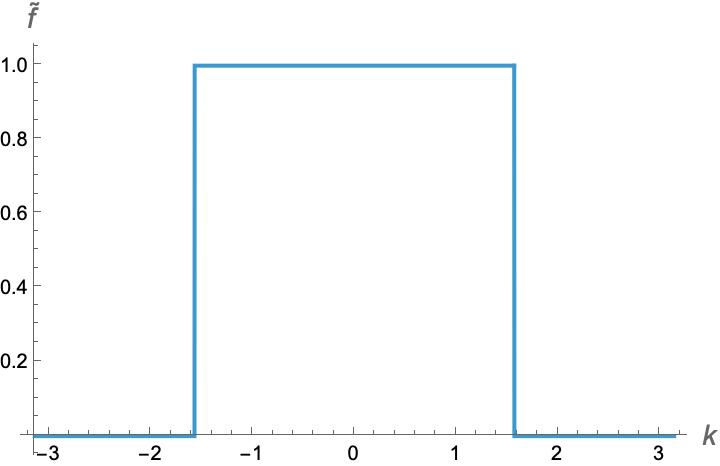}
\caption{The discrete Fourier transform $\tilde{f}(k)$ of the function $f(n)$ defining our family of Toeplitz matrices. (Note, these are dimensionless numbers.)}
\label{fig:ftilde}
\end{figure}

We can define a family of these Toeplitz matrices $\bar{D}$ as a function of the interaction range. Define a function $f(n)$ that takes an integer argument:
\begin{equation}
f(n) = \begin{cases} 1/2, & n=0 , \\ 0 & \text{even } n\ne0 , \\ (-1)^{(n-1)/2}/n\pi , & \text{odd } n .
\end{cases}
\end{equation} 
The matrix elements of $\bar{D}$ are $\bar{d}_{jj'} = f(j-j')$. We have not (yet) found a closed-form expression for the eigenvalues of this family of Toeplitz matrices, but we have found an analytical bound on the smallest eigenvalue of this matrix \cite{Parter61,Serra98}. First, we introduce the discrete Fourier transform of $f(n)$:
\begin{equation}
\tilde{f}(k) = \sum_{n=-\infty}^\infty e^{ink} f(n) ,\ \ \ k\in [-\pi,\pi] .
\end{equation} 
We can check that $\tilde{f}$ is convergent and real and has the properties $\tilde{f}(k) = \tilde{f}(-k)$ and $\tilde{f}(k + 2\pi) = \tilde{f}(k)$. Using the above observation connecting the matrix elements of $\bar{D}$ to the Taylor expansion of the arctangent, we can explicitly calculate
\begin{equation}
\tilde{f}(k) = \frac{1}{2} + \frac{1}{\pi}\left( \arctan(e^{ik}) + \arctan(e^{-ik}) \right) ,
\end{equation}
which produces a striking square wave, as shown in Figure~\ref{fig:ftilde}.
\vspace{-3pt}

The bound on the smallest eigenvalue follows from the following theorems:

\begin{Theorem}[\cite{Parter61,Serra98}]
If $\{A_M\}$ is a sequence  of $M\times M$ self-adjoint Toeplitz matrices (defined by a function $f(n)$ with well-defined transform $\tilde{f}(k)$), and if $\sigma_M$ is the minimum singular value of $A_M$, then $|\sigma_M - \sigma_\infty| < e^{-\alpha M}$ for some $\alpha>0$. (The smallest singular value of a self-adjoint matrix is the smallest absolute value of an eigenvalue, $|\lambda|$. Since the matrices $\{A_M\}$ are positive, this is the same as the smallest eigenvalue.)
\end{Theorem} 

\begin{Theorem}[\cite{Parter61,Serra98}]
The limit $\sigma_\infty$ of the smallest singular value of the matrix family $\{A_M\}$ is $\min_k \tilde{f}(k)$.
\end{Theorem}

As we can see in Figure~\ref{fig:ftilde}, the minimum value of $\tilde{f}(k)$ is 0. So the minimum eigenvalues of the matrices $\bar{D}$ approach 0 exponentially quickly as $M$ becomes large. (Note that the form of $\tilde{f}(k)$ also implies that the {\it maximum} eigenvalue approaches 1 exponentially quickly from below.) While we do not have an analytical expression for $\alpha$, we can estimate it numerically.

\subsubsection{Numerical Results}

\begin{figure}[H]
\includegraphics[width=3.8in]{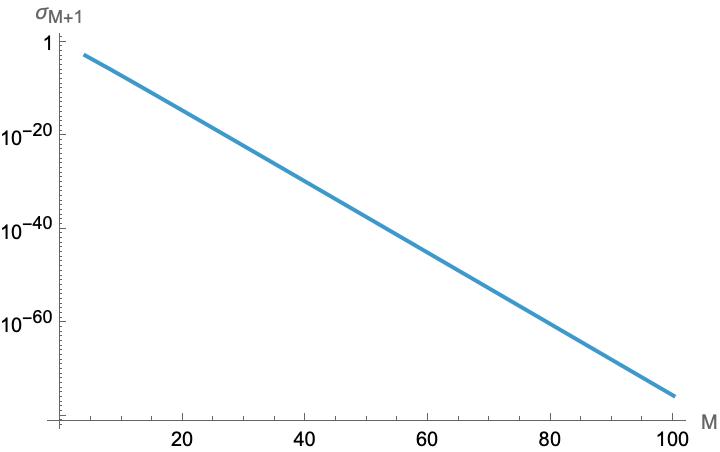}
\caption{The smallest eigenvalue of the $(M+1)\times(M+1)$ matrix $\bar{D}$ vs. the interaction range $M$. These values were calculated using high-precision numerics. (Note, these are \mbox{dimensionless numbers.)}}
\label{fig:minevals}
\end{figure}

We have numerically calculated the minimum eigenvalue for the $(M+1)\times(M+1)$ matrix $\bar{D}$ for an interaction range up to $M=100$. As shown in Figure~\ref{fig:minevals}, these eigenvalues drop exponentially with the range $M$. (Because these eigenvalues are so small, high-precision numerical calculations are required, with up to 110 digits of precision.) Numerical fitting suggests that these eigenvalues fall off as $e^{-\alpha M}$ for $\alpha \approx 1.755$. (We note that this is close to  $\sqrt\pi$, but in the absence of an analytical formula, this is likely to be a coincidence.)

The behavior of this eigenvalue tells us two things about locality and negative energies. First, the coupling to negative-energy modes is non-zero for every finite range. Any pair of coupled QCAs of this sort will have an amplitude to produce negative-energy bosons; to eliminate this amplitude requires non-vanishing interactions at arbitrary distances. But second, since the minimum eigenvalue falls off so rapidly with range, it is possible to suppress these amplitudes to the point where they are essentially negligible over any desired length of time. If we let the lattice size $N$ become very large, with a very small lattice spacing $\Delta x$, we can have the interaction range $M\Delta x$ be imperceptibly small---effectively indistinguishable from a point interaction at longer length scales. If we take the limit $\Delta x\rightarrow0$ and $N\rightarrow\infty$ to recover QFT, these become zero-range interactions.

\begin{figure}[H]
\includegraphics[width=3.8in]{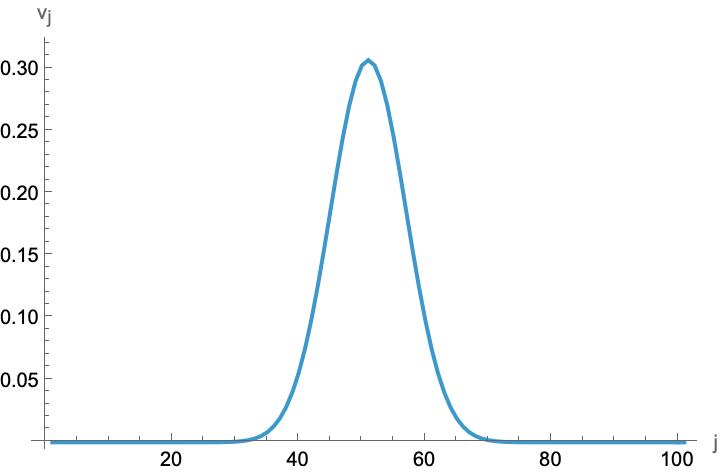}
\caption{The absolute value $|v_j|$ of the $j$th component of the eigenvector corresponding to the smallest eigenvalue of the $(M+1)\times(M+1)$ matrix $\bar{D}$ for $M=100$. Numerical fitting shows that it is extremely close to an exact Gaussian curve. (Note that, since this is an eigenvector, it can be multiplied by an arbitrary overall factor.)}
\label{fig:eigenvec101}
\end{figure}

In fact, this near-locality is even stronger than that. In Figure~\ref{fig:eigenvec101}, we plot the absolute value of the eigenvector component for the minimum value of $\bar{D}$ when $M=100$. To the eye, it appears to be a near-perfect Gaussian. Numerical fitting shows that it is, in fact, extremely close to a perfect Gaussian. This means that the amplitude to produce negative-energy bosons can be exponentially suppressed by an interaction over a finite range that also falls off rapidly away from the center.

One final note about the accuracy of these numerical calculations. The form of the matrix used to calculate the results shown in Figures~\ref{fig:minevals} and \ref{fig:eigenvec101} is given by the large $N$ approximation in Equation~(\ref{eq:realmatrix}) and is exact only in the limit $N\rightarrow\infty$. However, any corrections should scale like $O((M/N)^2)$ and so also be extremely small.

\section{Conclusions}\label{sec8}

The correspondence between quantum cellular automata (QCA) and quantum field theory (QFT) reveals a compelling perspective: that the fundamental laws of physics may emerge from an underlying information-processing algorithm. This \mbox{framework---where} QFTs arise from quantum computations on a discrete spacetime lattice---holds deep conceptual appeal. It provides a potential origin story for physical law itself. It offers a path toward addressing persistent challenges in continuum QFTs, such as renormalization, by embedding field dynamics in a fundamentally finite structure. It opens avenues for simulating QFTs on quantum hardware. And, most significantly, it raises the provocative possibility that the universe’s own fabric may be inherently discrete and computational.

We have shown in this paper that it is possible to construct a QCA that recovers free QED in the long-wavelength limit. In the case of bosons, this is accomplished by making the degrees of freedom of the local sites harmonic oscillator modes and coupling them by alternating quadratic interactions. This construction produces the Maxwell QFT by using a six-dimensional internal space. The result is a two-dimensional positive-energy sector (representing the two helicity states), a two-dimensional negative-energy counterpart, and a two-dimensional zero-energy sector that is static and decoupled from the rest.

Using oscillator modes makes the Hilbert space of the local sites infinite-dimensional. This is an interesting comparison with our earlier derivation of QCAs with fermionic statistics in two or more spatial dimensions, which required us to confine the state to the totally antisymmetric subspace of a QCA with high local dimension. It seems that to recover fermionic or bosonic statistics requires high-dimensional subsystems, albeit for different reasons.

To recover QFTs describing physical systems requires interactions. Such QCAs have been treated in a number of recent works \cite{Arrighi20,MlodinowBrun21,Eon23,Centofanti24}. One problem in adding interactions is the existence of negative-energy states. Negative-energy solutions for both the fermionic and bosonic cases follow from the time-reversal symmetry of the QCAs. This creates a dilemma, since a purely local interaction between two QCAs will then always be able to excite negative-energy bosons, which could produce a cascade of particle--anti-particle creation. We examined this problem in a 1D model and showed that amplitudes for negative-energy states cannot be eliminated by interactions with a finite range, but they can be suppressed without limit, with interactions that appear effectively local at longer length scales.

This demonstration that the amplitude of negative-energy solutions can be suppressed exponentially in the interaction range was shown for a particular form of interaction in a 1D model. Numerical studies are more challenging in higher spatial dimensions, since the size of the analogous matrix will grow like a higher power of the range. It will also no longer be a Toeplitz matrix, which may complicate the analysis. While we do not know, we expect that the suppression will most likely be even greater in higher dimensions, given the larger number of degrees of freedom available. It also raises the question of how zero-energy degrees of freedom, which arise in the bosonic QCA models derived above in higher dimensions, play a role in interacting theories. Could there also be other mechanisms to suppress the coupling to negative-energy (and zero-energy) modes, besides those discussed in this paper? Or could these modes end up playing a nontrivial role in a 3D theory?  These are subjects for future research.

\authorcontributions{Conceptualization, T.A.B. and L.M.; software, T.A.B.; validation, T.A.B. and L.M.; formal analysis, T.A.B. and L.M.; resources, T.A.B. and L.M.; data curation, N/A; writing, T.A.B. and L.M. contributed equally to all aspects of the writing, review and editing; visualization, T.A.B.; supervision, N/A; project administration, T.A.B.; funding acquisition, T.A.B.. All authors have read and agreed to the published version of the manuscript.} 

\funding{This work was supported in part by National Science Foundation Grant PHYS-2310794.}

\acknowledgments{TAB and LM gratefully acknowledge interesting conversations with Erhard Seiler, David Berenstein, Juan Garcia Nila, and Ammar Babar. We also appreciate the patience and support of Shirly Wu and the editorial staff at \textit{Entropy}.}

\conflictsofinterest{The authors declare no conflicts of interest.}

\begin{adjustwidth}{-\extralength}{0cm}
\reftitle{References}
\isAPAandChicago{}{%

}
\PublishersNote{}
\end{adjustwidth}
\end{document}